\documentclass[11pt]{article}

\usepackage[utf8]{inputenc}
\pdfoutput=1
\usepackage{jheppub_kim}

\usepackage{soul}
\usepackage[figtopcap]{subfigure}
\usepackage{amsmath,amssymb,dsfont}
\usepackage{hyperref}
\usepackage[T1]{fontenc} 
\usepackage{txfonts}
\usepackage{newlfont}
\usepackage{times}

\newcommand{\bq}{\begin{equation}}
 \newcommand{\eq}{\end{equation}}
 \newcommand{\bqn}{\begin{eqnarray}}
 \newcommand{\eqn}{\end{eqnarray}}
 \newcommand{\ba}{\begin{aligned}}
 \newcommand{\ea}{\end{aligned}}
 \newcommand{\nb}{\nonumber}
 \newcommand{\lb}{\label}

\newcommand\be{\begin{equation}}
\newcommand\ee{\end{equation}}
\newcommand\bea{\begin{eqnarray}}
\newcommand\eea{\end{eqnarray}}
\newcommand\bseq{\begin{subequations}} 
\newcommand\eseq{\end{subequations}}
\newcommand\bcas{\begin{cases}}
\newcommand\ecas{\end{cases}}

\title{Einstein-\AE ther Gravity in the light of Event
Horizon Telescope
Observations of
M87*}

\author[a]{Mohsen Khodadi}
\author[b,c,d]{and Emmanuel N. Saridakis}

\affiliation[a]{School of Astronomy, Institute for Research in Fundamental
Sciences
(IPM),\\ P.~O.~Box 19395-5531, Tehran, Iran}
 \affiliation[b]{National Observatory of Athens, Lofos Nymfon, 11852 Athens,
Greece}

  \affiliation[c]{CAS Key Laboratory for Researches in Galaxies and Cosmology,
Department of Astronomy, University of Science and Technology of China, Hefei,
Anhui 230026, P.R. China}
 \affiliation[d]{School of Astronomy, School of Physical Sciences,
University of Science and Technology of China, Hefei 230026, P.R. China}

\emailAdd{m.khodadi@ipm.ir}
\emailAdd{msaridak@noa.gr}

\abstract{We   investigate  Einstein-\AE ther  gravity  in light of
the  recent  Event Horizon Telescope (EHT) observations of the M87*.  The
shape and size of the observed black hole shadow contains information of the
geometry in its vicinity, and thus one can  consider it as a potential probe
to investigate   different   gravitational theories, since the involved
calculation framework  is  enriched with  different size-rotation features as
well as with extra model parameters. In the case of Einstein-\AE ther  gravity
the black hole solutions   include    two classes  depending on   the
involved   \ae{}ther parameters.  We
calculate the corresponding  photon   effective potential,  the unstable photon sphere radius, and finally   the induced
angular size, which combined with the    mass and  the distance can lead to a
single prediction
that quantifies the black hole shadow,   namely the diameter per  unit mass
$d$. Since  $d_{M87*}$ is
observationally known from the EHT Probe, we extract  the corresponding
parameter regions  in order to obtain consistency.
We find that Einstein-\AE ther black hole solutions    agree 
with the shadow size of EHT M87*, if 
the involved \AE ther parameters are restricted within specific ranges, along 
with an upper bound on the dimensionless spin parameter $a$, which is verified 
by a full scan of the parameter space within $1\sigma$-error.}

\keywords{M87* observations, Event Horizon Telescope, Einstein-\AE ther
gravity, Black holes}

\begin{document}

\maketitle

\section{Introduction}

In recent years, due  to considerable developments in observational astronomy,
accessible doors such as X-ray binaries, gravitational waves, and black-hole
shadow have been opened onto the study of black holes (BHs), leading to the
possibility to investigate the nature and traits of gravity in the strong-field
regime.
The first  are
actually X-rays emitted from a binary system due to matter  falling from one
normal star into its companion, which usually is a collapsed star, such as a
neutron star or a BH \cite{Shakura:1972te}. On the other hand, gravitational
waves observations arise  from a series of binary BH and neutron star mergers
recorded by LIGO-Virgo
Consortium \cite{Abbott:2016blz}-\cite{Monitor:2017mdv}. It is  expected that in
the not too distant future, by improving the signal-to-noise due to the
increased number of merging events,  such observations  will
be able to reveal more deep aspects of BH physics.

The black-hole  shadow  has become one of
the most exciting events in observational astronomy,
in the light of recording the first stunning new radio images of the
supermassive BH that resides  at the center of nearby galaxy M87* by
\textit{``Event
Horizon Telescope''} (EHT) (which incorporates the technology
called \emph{``mm-band Very Long Baseline Interferometry''} (VLBI) and images
the emissions close to supermassive BHs)
\cite{Akiyama:2019cqa}-\cite{Akiyama:2019eap} at April 2019.
In particular,  it provided the
first visual evidence indicating directly the existence of a compact object
 such as a supermassive BH.  Note that the defining
feature that distinguished a BH from other objects such as a wormhole or a
naked singularity is the event horizon, the boundary from within which nothing
can  escape. Due to the controversy about the distinction between a BH shadow
and a wormhole \cite{Ohgami:2015nra,Shaikh:2018kfv} or naked singularity
\cite{Virbhadra:2002ju,Virbhadra:2007kw,Ortiz:2015rma,Shaikh:2018lcc,Joshi:2020tlq} ones, as well as due to the
lack of recording thermal radiation indicating the direct existence of event
horizon \cite{Broderick:2009ph,Bambi:2012bh},  it is
reasonable to adopt a conservative approach and consider that the shadow image
released by the EHT collaboration  actually reveals an image   near to the
event horizon, at least as close as the light ring orbits
\cite{Cunha:2018acu}.

 Historically
the concept of the black hole shadow comes from the 70's with the seminal works
\cite{Synge:1966okc,Luminet:1979nyg} and \cite{Bardeen:1973}, which
respectively had been performed for the cases of   Schwarzschild BH and
rotating Kerr BH. In the light of these studies it was found that the
non-rotating
BH has a perfect circular shadow \cite{Narayan:2019imo},  while by taking the
rotation into account the shape of shadow is elongated due to the dragging
effect.
Nevertheless, the idea to image the shadow of BH  by taking advantage of
 VLBI technology,   appeared in \cite{Falcke:1999pj}.

The dark area over a brighter background in the center of the image  is termed
the \emph{``BH's shadow''}, and it is predicted by general relativity as the
null geodesics in the strong gravity area \cite{Cunha:2018acu,Chandra}.
Generally speaking, photons follow several paths depending on their angular
momentum: large, small, and critical. The first sets of photons coming from
infinity will be bounded back to infinity by the gravitational potential of the
BH. The second sets of photons will fall into the BH, which results in a dark
area for the distant observer. However, the third sets, namely the photons with
critical angular momentum, will swirl around the BH one ring by one ring which
makes the unstable boundary of the shadow.
This is the same as the shining halo
observed in EHT image, which is expected to be comprised of photons from the
hot, radiating gas that surrounds the BH, whose paths have been bent around the
BH before arriving at the telescopes. Although close to the BH shadow  the
strong gravitational lensing may be quite obvious, however, it is expected that
the shadows are far more easily to be observed, since unlike one-dimensional
lensing images   the BH shadows are two-dimensional dark regions.

Extracting the information saved in the shadow image released by EHT, which due
to the new imaging techniques enjoys ultra-high angular resolution,
conducts us  to understand in a more transparent way   significant issues such
as  the matter accretion process and BH jets. It also able to shed light on various metric theories of gravity \cite{Younsi:2016azx,Mizuno:2018lxz,DeFalco:2021klh}.
Hence, after the EHT
announcement, a large amount of research has been devoted    in calculating the
shadows of a vast class of BH
solutions
\cite{Shaikh:2019fpu}-\cite{Ghosh:2020spb} and the confrontation
with the extracted information from
EHT BH shadow image of M87* \cite{Davoudiasl:2019nlo}-\cite{Saurabh:2020zqg},
investigating   fundamental physics
issues too \cite{Bambi:2019tjh}-\cite{Psaltis:2020lvx}.
Additionally, one can also examine the BH shadow in the interplay with other
related concepts, such as quasinormal modes, deflection of light,
quasiperiodic oscillations, etc
\cite{Jusufi:2019ltj}-\cite{Ghasemi-Nodehi:2020oiz}. 
Moreover, some   BH shadow studies, in the light of employing 
general relativistic (GR) magnetohydrodynamic simulations of magnetized 
accretion flows onto BH, are able to provide a more solid knowledge, see 
e.g \cite{Mizuno:2018lxz,Olivares:2019dsc,White:2019wix,Nathanail:2020wap,
Bronzwaer:2020vix,Cruz-Osorio:2021gnz}.

Nevertheless, given that the shape and size of the shadow contains traces of the
geometry vicinity  of the BH, and similarly to the  quasinormal modes approach,
one can   consider  the shadow as a potential probe to investigate the BH
structure within different   gravitational theories.
In particular, although general relativity is a well-behaved theory at low
energy, infrared
(IR), scales and has come out proudly from numerous tests,  when high energy,
ultra-violet (UV) effects are involved one might need gravitational extensions
or modifications, namely   theories that possess general relativity as
a low-energy limit  but in general having a richer structure. Hence, one
expects advantages such as improvements of the
  singularity   behavior,  of renormalizability, or of cosmological
phenomenology.

One of the fundamental principles of general relativity  that might be altered
in the UV limit due to   the quantization of gravity is the
 Lorentz's invariance  (LI), since   Lorentz symmetry is not
fundamental in the sense that it depends on the scale we are exploring
nature's energy
\cite{Mattingly:2005re,Will:2005va,Liberati:2015dja}.
Breaking Lorentz symmetry in the UV
may imply the existence of a preferred frame at  Planck scale.
 Small departures from LI  could be used in order to  study
physics at the fundamental level, via investigation of  Standard Model
Extensions as an effective field theory framework
\cite{Colladay:1998fq} (see also \cite{Kostelecky:2003fs}).
By extending the LI breaking   into   gravity sector, a number of well known
modified theories arise,  such as Einstein-\AE ther
\cite{Jacobson:2000xp,Eling:2004dk,Jacobson:2008aj,Foster:2005dk,Elliott:2005va,
Li:2007vz,Yagi:2013ava},   Horava-Lifshitz
\cite{Horava:2009uw,Mukohyama:2010xz,Wang:2017brl},   mimetic gravity
\cite{Chamseddine:2013kea,Chamseddine:2014vna}, Finsler gravity
\cite{Basilakos:2013hua,Ikeda:2019ckp}
etc. Although the LI breaking in the above theories arise
from different mechanisms, one can explore the general features in a unified
way \cite{Sebastiani:2016ras}.

The Einstein-\AE ther (EA)  theory    is a generally covariant theory
of gravity, which violates the LI locally by possessing a dynamical,
unit-norm and timelike vector field, called   \textit{``\ae{}ther
field''}, which defines a preferred timelike direction at each
spacetime point.
  EA gravity   includes a number of coupling constants   called \ae{}ther
parameters, which are  tightly constrained   by
gravitational wave events such as \textit{GW170817} and \textit{GRB170817A}
\cite{TheLIGOScientific:2017qsa,Monitor:2017mdv,Oost:2018tcv}.
Similarly to most   modified theories of gravity, one can investigate BH
solutions in EA theory, considering an asymptotically flat
spacetime.
In particular, using two specific mixing of the \ae{}ther parameters, one
derives two sets of static and spherically symmetric
BH solutions \cite{Eling:2006ec,Barausse:2011pu}, while  extension to charged
solutions has been obtained   in \cite{Ding:2015kba} (see also
\cite{Zhang:2020too} for a systematic study of the spherically symmetric static
spacetimes in this framework).

In the present work we are interested in investigating EA theory using the
EHT observations of the M87*.
  In particular, since the EA-gravity BH solutions incorporate  the effects
of the \ae{}ther field,
this field will affect  the corresponding shadow too, and
hence   confrontation with  EHT observations will reveal if the theory at hand
is in agreement with them or if it must be constrained or excluded completely.
The crucial ingredient of the analysis is that although the
EHT's collaboration focuses on the Kerr BH solution in the background of
general relativity 
\footnote{Note that  in EHT papers \cite{Akiyama:2019cqa} and 
\cite{Akiyama:2019fyp} the possibility of   BH alternatives apart from Kerr was 
examined too. In general,  these alternatives are classified into three
principal classes: (i) BHs admitted by GR along with additional fields; (ii) BH 
solutions arisen from modified theories of gravity; (iii) BH mimickers, namely 
exotic compact objects that are allowed in GR or in modified gravities.} 
\cite{Akiyama:2019fyp,Psaltis:2014mca,Psaltis:2020lvx},
namely incorporating the angular size and rotation in a specific and relative
restricted framework, allowing for a modified gravity as the underlying theory
enriches the calculation framework with  different size-rotation features as
well as with extra model parameters. And this procedure will be more efficient
in theories which possess various corrections on the  Kerr-metric solutions,
such is the case in Einstein-\AE ther gravity.  Besides, since the validity
of the assumptions used in the measurement (for instance on rotation) cannot be
deduced solely from the correspondence between theory and data,    one can
have theories beyond general relativity that incorporate successfully  the data
 too \cite{Psaltis:2020lvx}.

 The manuscript  is structured as follows:  In
Section \ref{BH} we briefly  review   the slowly rotating BH solutions in
the EA gravity.
In Section  \ref{sh}  we determine the null geodesics equations as well
as the orbital equations of photons, and we investigate   the shadows related
to two types of slowly rotating  AE BH solutions.
In  Section  \ref{EHT} we confront the obtained shadows with the EHT image of
supermassive object in M87*, and we extract explicit constraints on the
coupling parameters of the EA theory. We eventually summarize our results and
conclude in  Section \ref{co}. Throughout the manuscript  we adopt natural
units where
$\hbar=k_B = c = 1$.

\section{Rotating black-hole solutions in Einstein-\AE ther  gravity}\label{BH}

In this Section we present the extraction of slowly rotating black-hole
solutions in   the framework of Einstein-\AE ther (AE)  gravity.
The action of the EA theory is \cite{Jacobson:2000xp,Jacobson:2008aj}
\bqn
S =S_{EH}+S_{AE}=\frac{1}{16 \pi G_{AE}} \int d^4 x \sqrt{-g} \Big(R+
\mathcal{L}_{AE}\Big),
\eqn
which includes the standard Einstein-Hilbert action $S_{EH}$ plus the \ae ther
action $S_{AE}$. In the above action, $R$, $G_{EA}$ and $\mathcal{L}_{AE}$ refer
respectively  to the Ricci scalar, the \AE ther gravitational constant and the
Lagrangian of the \AE ther field $u^\mu$, which is defined as
\bqn
\mathcal{L}_{AE} \equiv -\left(c_1 g^{\alpha \beta} g_{\mu\nu} + c_2
\delta^{\alpha}_{\mu} \delta ^{\beta}_{\nu} + c_3  \delta^{\alpha}_{\nu} \delta
^{\beta}_{\mu} - c_4 u^{\alpha} u^{\beta} g_{\mu\nu}\right) (\nabla_\alpha
u^\mu) (\nabla_{\beta} u^\nu) + \lambda_0 (u^2 +1)~.
\eqn
Here, $\lambda_0$ is a Lagrangian multiplier, ensuring that the \AE ther
four-velocity $u^{\alpha}$ is always timelike (i.e. $u^2=-1$).
All of four
coupling constants $(c_1,c_2,c_3,c_4)$ in the above expression are
dimensionless, and thus  $G_{AE}$ is linked to the Newtonian constant
$G_N$ via two of them, namely $G_{AE} = \frac{2G_N}{2-c_1-c_4}$
\cite{Carroll:2004ai}. These   coupling constants
  subject to    theoretical and observational
constraints such as
\cite{Jacobson:2007fh,Jacobson:2008aj,Berglund:2012bu}
\bqn
0\leq c_1+c_4<2,~~~2+c_1+c_3+3c_2>0,~~~c_1+c_3<1~.
\eqn
Variations of the total action with respect to $g_{\mu\nu}$, $u^{\alpha}$,
$\lambda_0$ yield, respectively, the field equations
\bqn
 &&R^{\mu\nu} - \frac{1}{2} g^{\mu\nu} R = 8 \pi G_{AE} T_{AE}^{\mu\nu},
\lb{einstein_equation}\\
&&\nabla_{\mu} J^{\mu}_{\;\;\;\alpha} + c_4 a_{\mu}\nabla_{\alpha}u^{\mu} +
\lambda_0 u_{\alpha}=0, \lb{aether_equation}\\
&&g_{\mu\nu} u^{\mu} u^{\nu} =-1 \lb{lambda_equation}~,
\eqn
where
\bqn
&&T^{\ae}_{\alpha\beta} \equiv
\nabla_{\mu}\Big(J^{\mu}_{\;\;\;(\alpha}u_{\beta)} +
J_{(\alpha\beta)}u^{\mu}-u_{(\beta}J_{\alpha)}^{\;\;\;\mu}\Big)
+ c_1\Big[\left(\nabla_{\alpha}u_{\mu}\right)\left(\nabla_{\beta}u^{\mu}\right)
-
\left(\nabla_{\mu}u_{\alpha}\right)\left(\nabla^{\mu}u_{\beta}\right)\Big]
\nb\\
&&\ \ \ \ \ \ \
 \ \ \ \
 + c_4 a_{\alpha}a_{\beta}    + \lambda  u_{\alpha}u_{\beta} -
\frac{1}{2}  g_{\alpha\beta} J^{\delta}_{\;\;\sigma}
\nabla_{\delta}u^{\sigma},\lb{current_equation}\\
&&J^{\alpha}_{\;\;\;\mu} \equiv \left(c_1 g^{\alpha \beta} g_{\mu\nu} + c_2
\delta^{\alpha}_{\mu} \delta ^{\beta}_{\nu} + c_3  \delta^{\alpha}_{\nu} \delta
^{\beta}_{\mu} - c_4 u^{\alpha} u^{\beta} g_{\mu\nu}\right)
\nabla_{\beta}u^{\nu},\lb{EM_equation}\\
&&a^{\mu} \equiv u^{\alpha}\nabla_{\alpha}u^{\mu}.
\eqn
From Eqs.(\ref{aether_equation}) and (\ref{lambda_equation}), we
straightforwardly find that
\bqn \lb{2.7}
\lambda_0 = u_{\beta}\nabla_{\alpha}J^{\alpha\beta} + c_4 a_{\lambda}a^{\lambda}.
\eqn
The vector $a^\mu$ in the above \AE ther stress tensor  denotes the
acceleration of the \AE ther field, and is orthogonal to \AE ther field itself
($u\cdot a=0$), and therefore        in its absence, i.e
$u^{\alpha}\nabla_{\alpha}u^{\mu}=0$,
  one can neglect the coupling constant $c_4$ in the field equations
\cite{Carroll:2004ai}.

 We proceed by  employing the Eddington-Finklestein coordinate system, which
respects static spherical symmetry. Then
the metric for EA black holes can be written as
\bqn \lb{metric_EF}
ds^2 = -e(r) dv^2 +2 f(r) dv dr + r^2 (d\theta^2+\sin^2\theta d\phi^2)~,
\eqn
with the   Killing vector $\chi^a=(1,0,0,0)$,
where $e(r)$, $f(r)$ are  $r$-dependent functions. Additionally,
we consider the    \ae ther vector parametrization
\bqn
  u^{a}(r) = (\alpha(r), \beta(r), 0, 0 )~,
\eqn
with    $\alpha(r)$, and $\beta(r)$ the involved  functions.
Concerning  the metric components at infinity (boundary conditions) we
require to correspond to  the asymptotically
flat solution, while those for the \ae{}ther components are set as $ u^a =
(1,0,0,0)$.

Under specific coupling constant $c_i$  choices,  for the static
spherically symmetric black hole
solutions in EA theory there can be two types of exact solutions
\cite{Eling:2006ec,Barausse:2011pu}.
By avoiding the details, the first solution corresponds to the special choice of
coupling constants
$c_{14}=0$ (where $c_{14} \equiv c_1+c_4$) and $c_{123} \neq 0$ (where $c_{123}
\equiv c_1+c_2+c_3$)
and takes the following form:
\bqn
\label{Sol1a}
&&e(r) = 1- \frac{2M}{r} - \frac{27 c_{13}}{256(1-c_{13})} \left(\frac{2M}{r}
\right)^4,\\
&& f(r) = 1, \lb{e14} \\
&&\alpha(r) = \left[ \frac{3 \sqrt{3}}{16\sqrt{1-c_{13}}}
\left(\frac{2M}{r}\right)^2 + \sqrt{1-\frac{2M}{r} + \frac{27}{256}
\left(\frac{2M}{r}\right)^4}\right]^{-1},\\
&&\beta(r) =- \frac{3 \sqrt{3}}{16\sqrt{1-c_{13}}}
\left(\frac{2M}{r}\right)^2,\label{Sol1d}
\eqn
while the second solution corresponds to $c_{123}=0$ and reads as
\bqn
\label{Sol2a}
&&e(r) = 1- \frac{2M}{r} -\frac{2c_{13} - c_{14}}{8(1-c_{13})}
\left(\frac{2M}{r} \right)^2,\\
&& f(r) = 1,  \lb{e123}\\
&&\alpha(r) =  \frac{1}{1+ \left(\sqrt{\frac{2-c_{14}}{2(1-c_{13})}} -1
\right)\frac{M}{r}},\\
&&
\beta(r) = - \sqrt{\frac{2-c_{14}}{8(1-c_{13})}} \frac{2M}{r}~,
\label{Sol2d}
\eqn
where $c_{13}\equiv c_1+c_3$.
It is clear that by fixing $c_{13}=0$ in the first solution and
$c_{13}=0=c_{14}$ in the second solution,
then we recover the standard Schwarzschild BH, as expected.

Finally, since usually the metric is written in the form of the $(t, \;
r,\; \theta, \; \phi)$ coordinates,
using the   coordinate transformation
$
dt = dv - \frac{dr }{e(r)}, \;\; dr=dr~,
$
the metric (\ref{metric_EF}) in the Eddington-Finklestein coordinate system, can
be re-expressed as
\bqn\lb{metric}
ds^2 = - e(r)dt^2+\frac{dr^2}{e(r)} +  r^2 (d\theta^2+\sin^2\theta d\phi^2)~,
\eqn
and hence   the \ae{}ther field becomes
\bqn
u^a= \left(\alpha(r) - \frac{\beta(r)}{e(r)}, \beta(r), 0, 0\right).
\eqn

The main difficulty in
deriving BH solutions in Lorentz violating theories is related to the existence
of casual boundaries, indicating an event horizon, as an essential requirement
for BHs, \cite{Berglund:2012bu,Gorji:2019rlm}.
Thus, in EA-gravity
(as well as in Horava-Lifshitz one), due to complexities, we still do not have
 the fully rotating BH solution.
Despite this, one can utilize   spherically symmetric BH solutions  in the
Hartle-Thorne slow-rotation approximation
(first order), in order to derive the rotating BH solutions in the slow
limit.
 Hence,   applying  the well-known Hartle-Thorne metric
\cite{Hartle:1968si}
\bqn
ds^2 = -e(r) dt^2 + \frac{B(r) dr^2}{e(r)} + r^2 (d\theta^2+\sin^2\theta
d\phi^2)-\epsilon r^2
\sin^2 \theta\Omega(r, \theta) dt d\phi + \mathcal{O}(\epsilon^2)~,
\eqn
with $\epsilon$   representing a perturbative slow rotation parameter,
one can derive the rotating black hole solution in the slow rotation limit for
EA theory.
Here, the functions $e(r)$ and $B(r)$  denote the ``seed'' static,
spherically-symmetric
solutions when the frame dragging parameter   equals zero, namely
$\Omega(r, \theta) =0$. Moreover, the function $B(r)$
is usually fixed to unity, and the \ae{}ther configuration in
the slow-rotation
limit acquires the following form
\cite{Barausse:2015frm}
\bqn
u_a dx^a =\big(\beta(r)-e(r) \alpha(r)\big)dt +\frac{\beta(r)}{e(r)} dr
+ \epsilon \big(\beta(r)-e(r) \alpha(r)\big) \lambda(r, \theta) \sin^2 \theta d
\phi + \mathcal{O}(\epsilon^2)~,
\eqn
where the parameter $ \lambda(r, \theta)$ is connected to the \ae{}ther's
angular momentum per unit energy via
relation $ \lambda(r, \theta)=\frac{u_{\phi}}{u_t \sin^2\theta}$.

In order to satisfy the
asymptotically flat boundary conditions,   the functions $\Omega(r,\theta)$
and $\lambda(r, \theta)$ are required to
be $\theta$-independent, namely $\Omega(r, \theta)=\Omega(r)$ and $\lambda(r,
\theta) = \lambda(r)$ \cite{Barausse:2013nwa}.
For the first static solution (\ref{Sol1a})-(\ref{Sol1d}) there exists a
corresponding slowly rotating black hole solution,
with a spherically symmetric \ae ther field configuration ($\lambda(r)=0$) and
thus
\bqn \lb{om1}
\Omega(r)=\Omega_0+ \frac{4 J}{r^3}\,,
\eqn
namely
\bqn\lb{slow1}
ds^2 = -e(r) dt^2 + \frac{dr^2}{e(r)} + r^2 (d\theta^2+\sin^2\theta
d\phi^2)-\frac{4 M a}{r} \sin^2
\theta dt d\phi + \mathcal{O}(\epsilon^2)~,
\eqn
with $e(r)$ given by  (\ref{Sol1a}) and $\Omega_0$ an
integration constant that can be set to zero.
Note that for convenience we have replaced the angular momentum $J$ by
introducing the rotation parameter $a$ through \cite{Bambi:2019tjh}:
\bqn\lb{adeff}
 a\equiv\frac{J}{M}.
\eqn
Nevertheless, for the second static solution (\ref{Sol2a})-(\ref{Sol2d}) which
is obtained for $c_{123}=0$, one cannot find a closed form for the expression of
$\Omega(r)$ and $\lambda(r)$ except in the limit $c_\omega = c_1-c_3 \to
\infty$. Only in this  limit   the frame dragging potential $\Omega(r)$
becomes the same as in (\ref{om1})  \cite{Barausse:2015frm}.
 Therefore,  the metric of the second  rotating solution acquires the same
form   (\ref{slow1}),  but with $e(r)$ given by  (\ref{Sol2a}).
In summary, both underlying static spherically symmetry
solutions result in a unified slowly rotating metric. In the following  we
name the first and second
solutions respectively as Einstein-\AE ther I and II types of BH solutions.

We close this section by
 mentioning that EA theory in the limit $(c_1-c_3) \to \infty$
coincides to the non-projectable
Horava-Lifshitz  gravity in the IR limit, indicating that   solutions
in the EA theory with $(c_1-c_3) \to \infty$
are solutions of the Horava-Lifshitz   gravity, too \cite{Wang:2012nv}.
As a cross check it should be emphasized that
the parameters $(c_1, c_2, c_3, c_4)$ used in the present work are actually
connected to the parameters $(c_\theta, c_\sigma, c_\omega, c_a)$ in
\cite{Barausse:2015frm} through the relations $c_{\theta} = c_1+3c_2+c_3$,
$c_{\sigma } = c_1+c_3 = c_{13}$, $c_\omega = c_1- c_3$, and $c_a = c_1+c_4 =
c_{14}$.

\section{Black hole  shadow in Einstein-\AE ther gravity} \label{sh}

In this section we calculate the black hole  shadow profile in the framework of
Einstein-\AE ther gravity. If we have a BH solution, a
  dark shadow  is an observer-independent observable, which generally can be
defined
as an absorption cross-section of the gravitationally captured photon region
enclosed
by the innermost unstable photon orbits around a BH
\cite{Cunha:2018acu,Chandra}.
Hence,  the shadow is actually the border area between the captured photon
orbits
and the scattered photon orbits. Therefore,
in order to reveal the dark shadow
of the slowly rotating BHs  of  EA gravity, we have to investigate
the structure
of the photon geodesics. Additionally, by neglecting the interaction
between the electromagnetic sector and the \ae{}ther field, we still let
photons to follow null geodesics.

As usual, in order to study the geodesics structure of the photon trajectories,
we begin with the Hamilton-Jacobi equation
\bqn\lb{HJ}
\frac{\partial S}{\partial \lambda} = - \frac{1}{2}g^{\mu\nu} \frac{\partial
S}{\partial x^\mu} \frac{\partial S}{\partial x^\nu}~,
\eqn
where $S$ and $\lambda$ denote respectively the Jacobi action of the particle
(here photon) moving in the black hole spacetime,
and the affine parameter of the null geodesic. Concerning the massless photon
propagating on the null geodesics, the Jacobi action
$S$ can be separated as
\bqn\lb{Jacobi_action}
S = - E t + J \phi + S_{r}(r) + S_\theta (\theta)~,
\eqn
where $E$ and $J$ respectively address the energy and angular momentum of the
photon in the direction of
the rotation axis. Furthermore, the functions $S_r(r)$ and $S_\theta(\theta)$
have only  $r$ and $\theta$ dependencies, respectively.

By inserting the Jacobi action (\ref{Jacobi_action}) into the Hamilton-Jacobi
equation (\ref{HJ}),   using also
the metric components   (\ref{slow1}), we acquire
\bqn
e(r) r^2 \left(\frac{dS_r}{dr}\right)^2 +
\left(\frac{dS_\theta}{d\theta}\right)^2-\frac{E^2 r^2}{e(r)}+\frac{J^2}{\sin^2
\theta}
+\frac{4 E J M a}{r e(r)}=0~,
\eqn
where the neglected  terms are of $\mathcal{O}(a^2)$ due to their negligible
contribution in the slowly rotating limit ($a\ll1$).
Introducing a separation constant
 $\mathcal{K}$  we can separate  the above equation  as
\bqn
&&J^2 \cot ^2\theta+ \left(\frac{dS_\theta}{d\theta}\right)^2 = \mathcal{K},\\
&& r^2 e(r) \left(\frac{dS_r}{dr}\right)^2 - \frac{E^2 r^2}{e(r)} + \frac{4 E J
M a}{r e(r)} = - \mathcal{K}-J^2,
\eqn
 and through integration we respectively arrive at the solutions of
$S_\theta(r)$ and $S_r(r)$, namely
\begin{eqnarray}
&&S_{\theta} (\theta) = \int^\theta \sqrt{\Theta(\theta)} d\theta,\\
&&S_r(r) = \int^r \frac{\sqrt{R(r)}}{r^2 e(r)} dr,
\end{eqnarray}
where
\begin{eqnarray}
&&R(r) =  E^2r^4 - (\mathcal{K}+J^2) r^2 e(r) - 4 M a EJ r~,\\
&&\Theta(\theta) = \mathcal{K} - J^2 \cot ^2\theta.
\end{eqnarray}
Thus, the photon propagation obeys   the following four equations of
motion,  obtained from the variation of the Jacobi action with respect to the
affine parameter $\lambda$:
\begin{eqnarray}
&&\frac{dt}{d\lambda} = \frac{E}{e(r)} - \frac{2 M J a}{e(r)r^3},\\
&&
\frac{dr}{d\lambda} =\frac{\sqrt{R(r)}}{r^2}, \\
&&
\frac{d\theta}{d\lambda} =
\frac{\sqrt{\Theta(\theta)}}{r^2},\\
&&
\frac{d \phi}{d \lambda} = \frac{J }{\sin^2\theta r^2} - \frac{2 M E a}{r^3
e(r)}.
\end{eqnarray}

In order to investigate the photon trajectories   one usually expresses the
radial geodesics
in terms of the effective potential $U_{eff}(r)$ as
\bqn
 \left(\frac{dr}{d\lambda}\right)^2 + U_{eff} (r)= 0
\eqn
with
\bqn
U_{eff}(r) = -1 + \frac{e(r)}{r^2} (\xi^2 +\eta) + \frac{4 M  \eta
a}{r^3},
\label{Effpot1}
\eqn
where $\xi = \frac{J}{E},\;\;\eta = \frac{\mathcal{K}}{E^2}$.
The above two impact parameters $\xi$ and $\eta$ are actually the principle
quantities for determining the photon motion.
To obtain the geometric shape of the BH shadow, conventionally we have to find
the photon
critical circular orbit.
This can be extracted from the following unstable conditions:
\bqn\lb{condition}
U_{eff}(r)=0~,\;\; \frac{dU_{eff}(r)}{dr} =0 ~,\;\;\; \frac{d^2
U_{eff}(r)}{dr^2} <0~.
\eqn
We can extract  the geometric shape of the shadow via the
allowed
values of $\xi$ and $\eta$ that satisfy the above   conditions.
Thus,  with the implementation of  (\ref{condition}) we arrive  at
\bqn\lb{PSradius}
2(\eta+\xi^2)r e(r)-(\eta+\xi^2)r^2e'(r)+12M \xi a=0~.
\eqn
By solving this equation one   acquires the radius $r_{ps}$ of the
photon
sphere, which
since we have taken the rotation  effect   into account is expected to be
between the two values $r_{ps}^{\mp}$.

For slowly rotating BHs, solving   conditions (\ref{condition})  we
immediately find  that for the
spherical-orbit photon motion  the two parameters $\xi$ and $\eta$ have the
form
\bqn
\xi(r) &=& \frac{r^3 [r e'(r) - 2\,e(r)]}{4 M a [e(r)+r e'(r)]} ~, \\
\eta(r) &=& \frac{-r^6[-2e(r)+re'(r)]^2+48M^2
a^2r^2\big[e(r)+re'(r)\big]}{16M^2
a^2\big[e(r)+e'(r)\big]^2}~.
\eqn
Overall, the gravitational lensing effects result in deflection of the photon
passing a BH.
Specifically, some photons have the chance of reaching the distant
observer after being deflected by the BH, while some others will fall into it.
As a result,
the photons that cannot escape the black hole are the ones that create the
black hole shadow. As usual, to describe the shadow as seen by a distant
observer,  one introduces
the following two celestial coordinates $X$ and $Y$ \cite{Chandra}:
\bqn
X &=& \lim_{r_* \to \infty} \left(- r_*^2 \sin \theta_0
\frac{d\phi}{dr}\right),
\\
Y &=& \lim_{r_* \to \infty} r_*^2 \frac{d\theta}{dr},
\eqn
where $r_*$ and $\theta_0$  are respectively the distance between the observer
and the black hole, and the inclination angle between the line of sight of the
observer and the rotational axis of the black hole. By applying the geodesics
equations along with  the expressions $\frac{d\phi}{dr} =
\frac{d\phi/d\lambda}{dr/d\lambda}$ and $ \frac{d\theta}{dr} =
\frac{d\theta/d\lambda}{dr/d\lambda}$
we obtain
\begin{eqnarray}
 &&X = - \xi(r_{ps}) \csc\theta_0~, \\
 &&
 Y =  \sqrt{\eta (r_{ps}) -
\xi^2(r_{ps}) \cot^2\theta_0}~,
\end{eqnarray}
and therefore   these two celestial coordinates fulfill
\bqn
\label{fordelta}
X^2+Y^2 = \xi^2(r_{ps}) +\eta_{\bold ps}(r_{ps})~,
\eqn
where $r_{ps}$ is the aforementioned radius of the unstable photon sphere.
This is the expression of  the EA BHs shadow  in the slow rotation
limit.
As a self-consistency check, we can see  that in the
Schwarzschild limit, namely   $a\longrightarrow0$, the above equation becomes
$X^2+Y^2 =27 M^2$, as expected.

In the following subsections, by keeping   the leading order of the small
rotation
parameter $a\ll1$,   we will investigate  the shadow of
Einstein-\AE thertypes I and II BH solutions, separately.

\subsection{Einstein-\AE ther type I black hole solution}

\begin{figure}[ht]
\includegraphics[width=8cm,height=6cm]{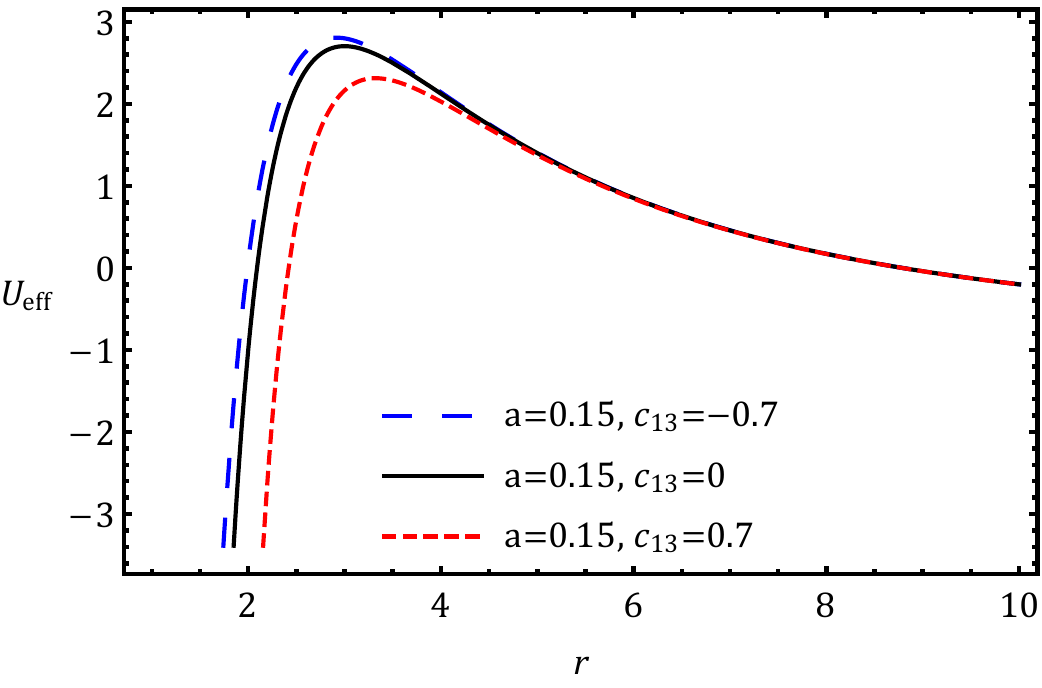}~~~~~
\includegraphics[width=6.5cm,height=6.5cm]{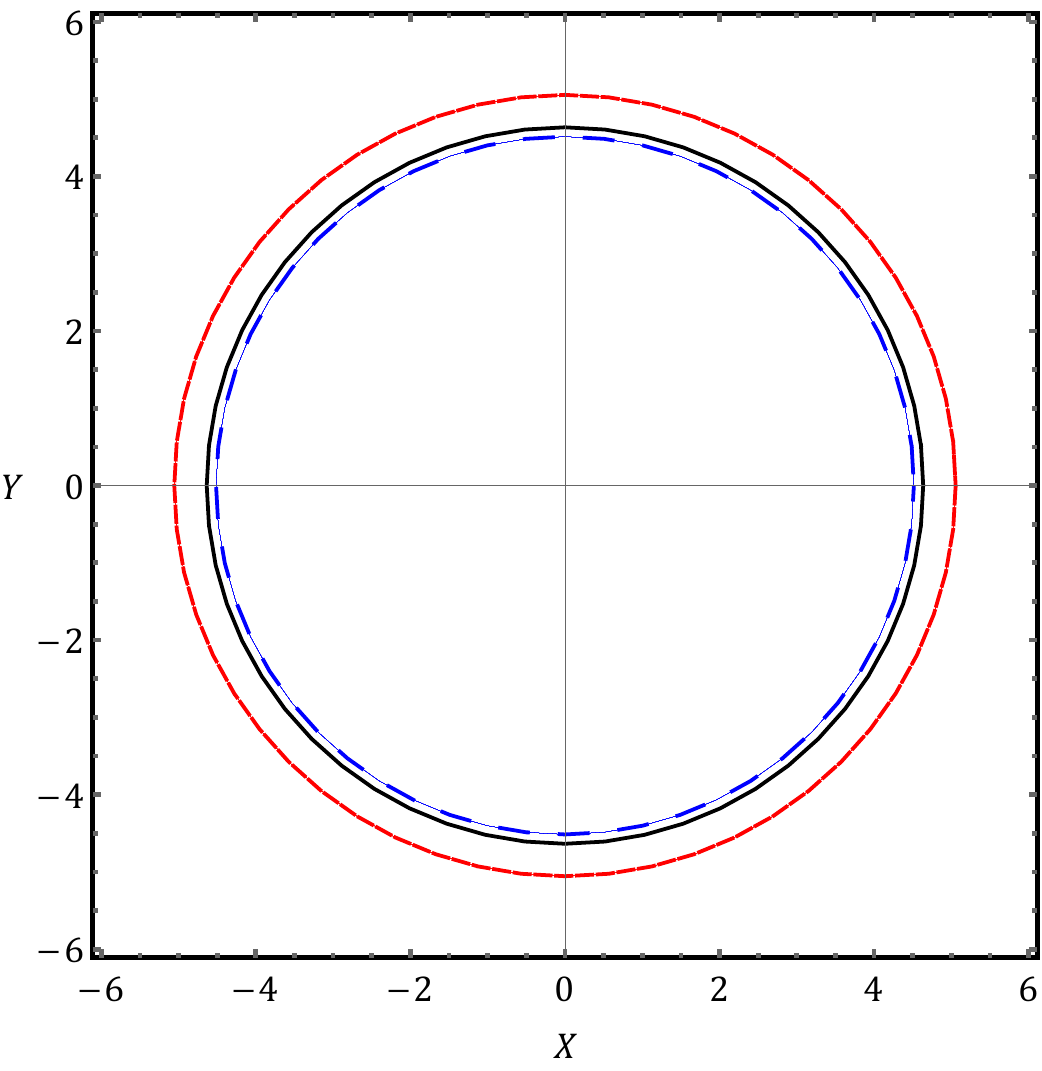}
	\caption{{\it{
	{\bf {Left graph:}} The effective potential of the
photons $U_{eff}$ as a function of the radial distance $r$, for the
Einstein-\AE ther type I black hole solution (\ref{slow1}) with (\ref{Sol1a}),
for various   choices of the  \ae{}ther parameter $c_{13}$  and the rotation
parameter $a$.
 {\bf {Right graph:}}
 The corresponding black hole shadow region according to (\ref{fordelta}). In
both graphs the $c_{13}$ values    have been un-realistically magnified in
order to be able to discriminate clearly its effects.
 }}}
	\label{Veff1}
\end{figure}

In this subsection we focus on the examination of the shadow features of the
Einstein-\AE ther type I black hole solution (\ref{slow1}),    with $e(r)$
given by  (\ref{Sol1a}). In the left panel of Fig.~\ref{Veff1}, 
by resrticting to the equatorial plane ($\theta=\pi/2$), we
present the effective potential $U_{eff}(r)$ given in  (\ref{Effpot1}),
obeyed by the photons in the background of the Einstein-\AE  ther type I BH
(i.e. when $c_{14}=0$ but $c_{123} \neq 0$) and . As we can see,
  $U_{eff}$ admits a
unique maximum, which reveals the presence of an unstable circular
orbit around what we expect for the Schwarzschild case, namely $r_{ps}\simeq
3$. Such an unstable
circular orbit implies that under any external
perturbation   the photons  will leave the
circular orbit. Instead these photons form a photon
sphere which is observable as a BH shadow
in the distant observer's frame. Additionally,
increasing the value of $c_{13}$ \ae ther parameter from   negative to
positive,   the peak of the curve shifts to lower values.
However, in the asymptotic $r$ regime, independently of the  $c_{13}$  values,
$U_{eff}$ tends to   $-1$,
which implies that  the photon motion  remains stable at infinity due to
its constant energy.

In order to  draw the shadow of the   BH solution we need to calculate
two essential
quantities, namely the event horizon radius $r_{e}$
and the radius of the unstable photon sphere $r_{ps}$. In order to find the
even horizon of metric (\ref{Sol1a}) we
have to solve
\bqn
r^4 - 2 M r^3 - \frac{27 c_{13}}{16(1-c_{13})} M^4=0~,
\eqn
which leads to
\begin{eqnarray}
\lb{ev}
&&r_{e_{1,2}}=\frac{3M^2}{2}-s\pm\frac{1}{2}\sqrt{-4s^2+3M^2-\frac{M^2}{s}},\\
&&r_{e_{3,
4 }}
=\frac{3M^2}{2}+s\pm\frac{1}{2}\sqrt{-4s^2+3M^2+\frac{M^2}{s}},
\end{eqnarray}
 where
\begin{eqnarray}
&&s=\sqrt{\frac{3M^2Q+Q^2+\Delta_0}{12Q}}, ~~~Q=\left(\frac{\Delta_1+
	\sqrt{\Delta_1^2-4\Delta_0^3}}{2}\right)^{1/3},\\
&&
\Delta_0=- \frac{81c_{13}}{4(1-c_{13})} M^4,~~~\Delta_1= \frac{729
c_{13}}{2(1-c_{13})} M^7  .
\end{eqnarray}
Solutions $r_{1,2}$ are imaginary and thus not physically interesting.
Nevertheless,     by setting $c_{13}$ to zero
 the third solution becomes
$r_3=2M$, as expected from Schwarzschild background. Thus, we deduce that
  $r_3$ addresses the event horizon
radius $r_e$ of the Einstein-\AE ther type I BH solution.

We proceed  by setting $\eta=0$, and inserting the $\xi_{ps}$ from $U_{eff}=0$
\bqn\lb{xi}
\xi_{ps}=\frac{2M a}{e(r_{ps})
r_{ps}}\left[-1+\sqrt{1+\frac{r_{ps}^4e(r_{ps})}{4M^2 a^2 }}\right]~,
\eqn
into (\ref{PSradius}). Then, using $e(r)$ from Eq.~(\ref{Sol1a}) we
finally result to the following equation
\begin{equation}
\lb{rps1}
 a^2\left(\alpha _1-8\right) \left(c_{13}-1\right) r_{ps}^3 \left[81
\alpha _1 c_{13}+16 \left(\alpha _1+16\right) \left(c_{13}-1\right) r_{ps}^4-48
	\left(\alpha _1+8\right) \left(c_{13}-1\right) r_{ps}^3\right]  =0~,
\end{equation}
with
\bqn\lb{rps2}
\alpha_1=\sqrt{\frac{64a^2\left(c_{13}-1\right) (r_{ps}-2) r_{ps}^3+16
\left(c_{13}-1\right)+27 c_{13}}{(c_{13}-1)a^2}}~.
\eqn
The solution of Eq. (\ref{rps1}) will provide   the radius of the photon
sphere
$r_{ps}$.
\begin{figure}[ht]
\begin{center}
\includegraphics[width=8cm,height=6cm]{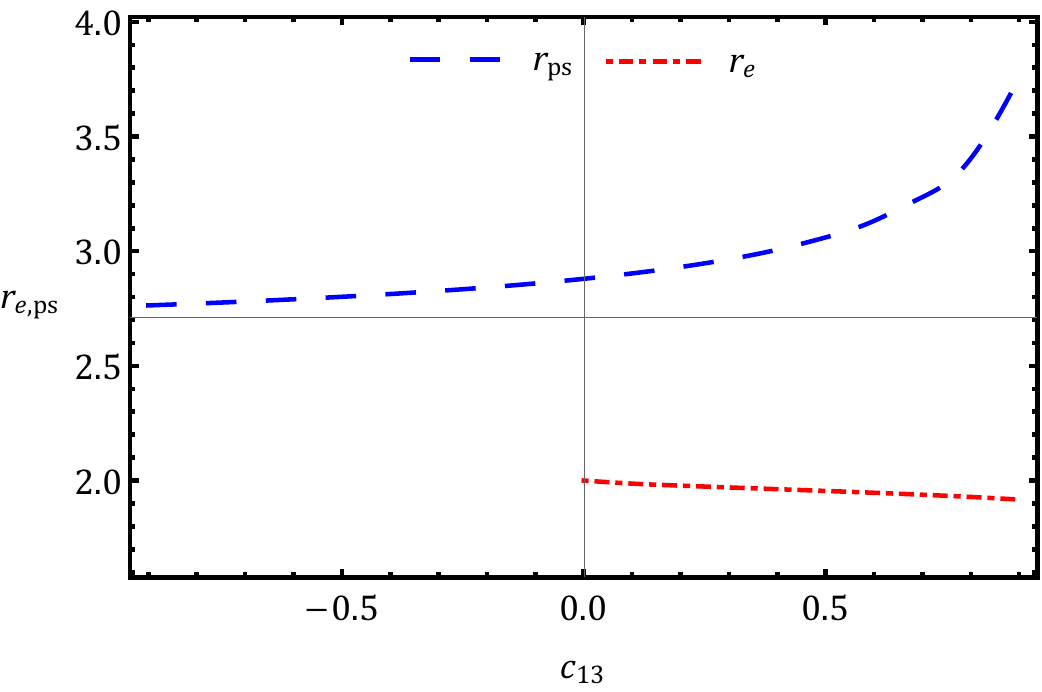}
	\caption{{\it{ The event horizon radius $r_{e}$
and the radius of the unstable photon sphere $r_{ps}$, in terms of the \ae{}ther
parameter $c_{13}$, for the Einstein-\AE
ther type I black hole solution (\ref{slow1}) with (\ref{Sol1a}),  for the
representative value $a=0.1$, as they arise by the numerical solution of
(\ref{ev}) and (\ref{rps1}) respectively. The interval of the $c_{13}$ values
 has been un-realistically magnified in
order to be able to discriminate clearly its effect. 
}}}
	\label{rerps}
\end{center}
\end{figure}

A first observation is that  in the limit $a\rightarrow0$  Eq.
(\ref{rps1}) reduces
to $e'(r_{ps})r_{ps}-2e(r_{ps})=0$, as expected \cite{Zhu:2019ura}. However, in
the general case   Eqs. (\ref{ev}) and (\ref{rps1}) cannot be solved
analytically, and therefore we elaborate them numerically and in Fig.
 \ref{rerps}  we depict the obtained solutions in terms of the
\ae{}ther parameter $c_{13}$.
As one can see, for $c_{13}<0$   the Einstein-\AE ther type I
solution becomes a naked singularity with  no   event
horizon.
However,  for non-negative $c_{13}$  the photon
sphere radius increases, in agreement with the corresponding shift
of the peak
of $U_{eff}$ in the left panel
of Fig.~\ref{Veff1}. Having in mind the
 general intuition that  for the formation of the
shadow the existence of the photon
sphere is essential (see however   \cite{Joshi:2020tlq} which shows that
for a  naked singularity there is the possibility of shadow formation  without a
photon sphere), one could expect that the BH shadow size   would be larger
than the shadow size of a naked singularity, as depicted in the
right panel of Fig.~\ref{Veff1}.
This feature might serve as a phenomenological signal for
distinguishing the BH from a naked singularity.

\subsection{Einstein-\AE ther type II black hole solution}

\begin{figure}[ht]
	\includegraphics[width=8cm,height=6cm]{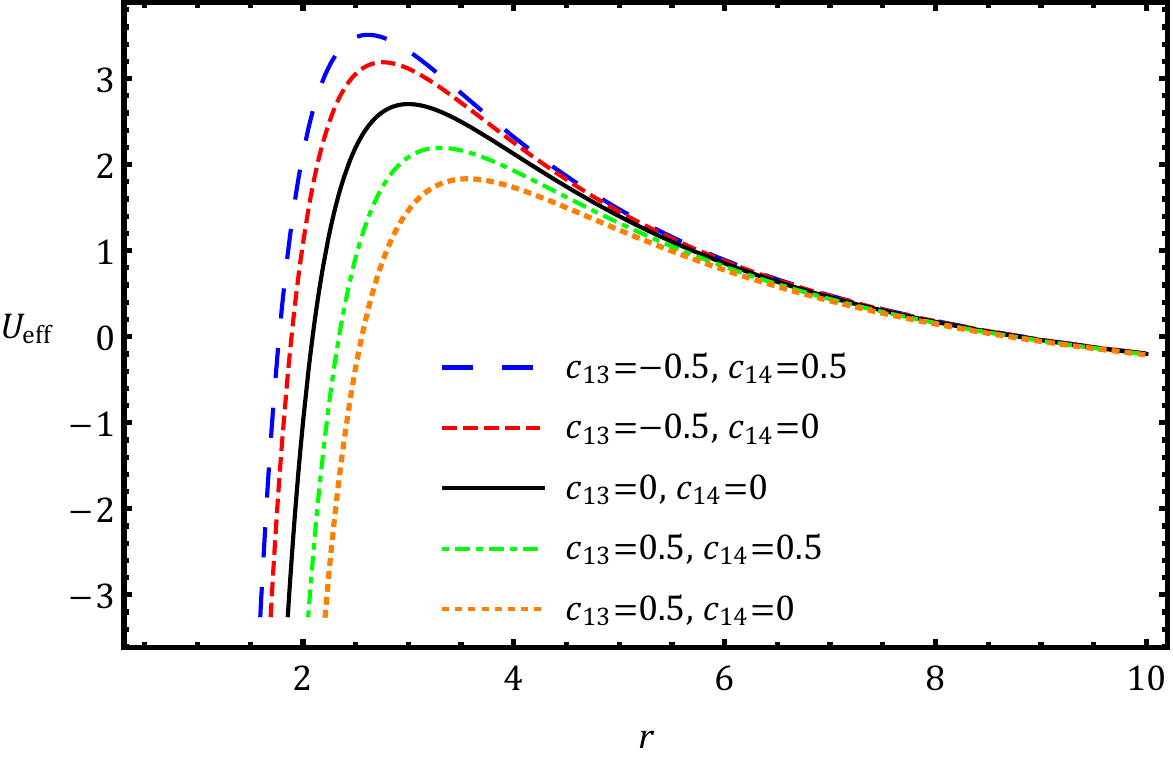}~~~
	\includegraphics[width=6.5cm,height=6.5cm]{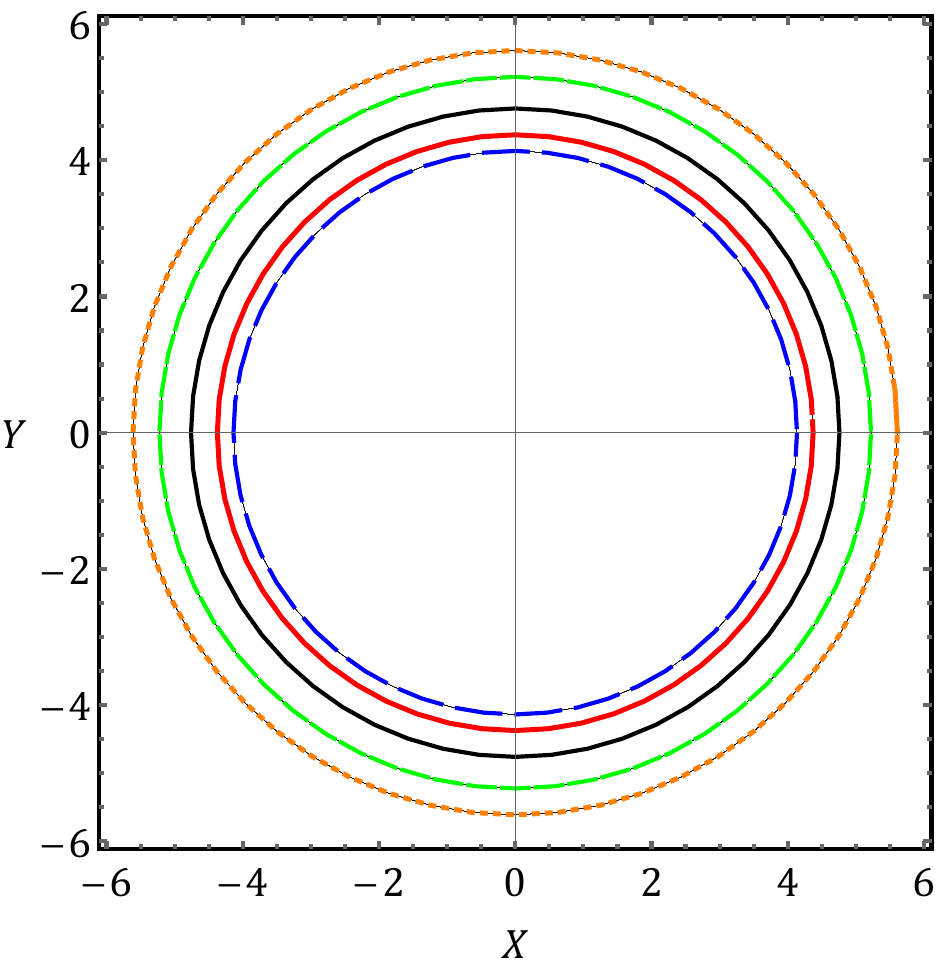}\\
	\caption{{\it{ {\bf {Left graph:}} The effective potential of the
photons $U_{eff}$ as a function of the radial distance $r$, for the
Einstein-\AE
ther type II black hole solution (\ref{slow1}) with (\ref{Sol2a}), for various
 choices of the
\ae{}ther parameters $c_{13}$ and $c_{14}$  and for rotation parameter $a=0.15$.
 {\bf {Right graph:}}
 The corresponding black hole shadow region according to (\ref{fordelta}).  In
both graphs the $c_{13}$ and $c_{14}$ values    have been un-realistically
magnified in
order to be able to discriminate clearly their effects. 
 }}}
	\label{Veff2}
\end{figure}

In this subsection we perform our analysis  for  the
Einstein-\AE ther type II black hole solution (\ref{slow1}),    with $e(r)$
given by  (\ref{Sol2a}). In the left panel of Fig.~\ref{Veff2},
  we   depict the effective potential
  $U_{eff}(r)$ given in  (\ref{Effpot1}), corresponding to  the
Einstein-\AE  ther type II BH ($c_{13}\neq0\neq c_{14}$ but with $c_{123}=0$).
As we can see,  the behavior of the curves
depends on the   \ae{}ther parameters values. In particular,
 moving from negative to positive values of $c_{13}$
the peak of $U_{eff}$ forms in larger distances with lower height,
however, as $c_{14}$  grows (within the allowed interval) the peak moves to
lower
distances with higher height. Thus, the photon  sphere radius will also be
different in the various cases.

Similarly to the previous solution subclass above, we can now proceed to the
investigation of the resulting  event horizon
radius    $r_e$ and of  the radius of the unstable photon sphere $r_{ps}$.
Using  the metric function (\ref{Sol2a}), we  find    $r_e$ by solving
\bqn
r^2- 2M r -\frac{(2c_{13} - c_{14})M^2}{2(1-c_{13})}=0~,
\eqn
and thus we obtain the two solutions
\bqn
r_{e\mp}=M\bigg(1\mp\sqrt{\frac{c_{14}-2}{2c_{13}-2}}\bigg)~.
\eqn
Amongst them, solution   $r_{e+}$ is the physically interesting one, since by
setting $c_{13}=0=c_{14}$ this is the one that
recovers the general relativity, Schwarzschild result, namely $r_e=2M$.
Additionally, as one can see,   for the  allowed values $0<c_{14}<2$ and
$c_{13}<1$ the BH has always a real event horizon radius and thus a naked
singularities cannot appear, unlike the previous Einstein-\AE ther type I
case.

We proceed by extracting
the photon sphere radius. Inserting (\ref{xi}) into
(\ref{PSradius}) and replacing $e(r)$ from  (\ref{Sol2a}) we acquire
\begin{equation}
 \lb{n1}
 \frac{a^2 \left(c_{13}-1\right) r_{ps}}{\alpha_3{}^2}\left\{\left(4-4 \alpha
_2\right){}^2 \left[2 c_{13} (1-r_{ps})+2
r_{ps}-c_{14}\right]-\alpha_3\left(4-4
\alpha _2\right){}^2
-12 \alpha_3\left(4 \alpha _2-4\right)\right\} =0  ,
\end{equation}
where
\bqn
&&\alpha_2=\sqrt{\frac{r_{ps}^2 \alpha_3}{8 a^2
\left(c_{13}-1\right)}+1}~,\nonumber\\
&&
\alpha_3=-2 c_{13} (r_{ps}-1)^2+2r (r_{ps}-2)+c_{14}~.
\eqn
Similarly to the Einstein-\AE ther type I
case,   in the limit $a\rightarrow0$  Eq. (\ref{n1}) also
reduces to   $e'(r_{ps})r_{ps}-2e(r_{ps})=0$, as
expected \cite{Zhu:2019ura}.

Since (\ref{n1}) cannot be solved analytically, we elaborate it numerically. In
the right graph of Fig.~\ref{Veff2}
we depict the corresponding BH shadow for the same parameter choices of the
left graph.
  As we observe,  as the peak of $U_{eff}$ shifts towards smaller
(larger) values, the
shadow size   decreases (increases).

\section{The \ae{}ther parameters $c_{13}$ and $c_{14}$ and M87* observations
}\label{EHT}

We can now proceed to the investigation of the constraints on the  \ae{}ther
parameters $c_{13}$ and $c_{14}$  that arise from the EHT
Observations of the shadow of M87*. We will focus on the two classes of
solutions, namely
Einstein-\AE ther  types I and II BH solutions.  In particular,
observing  Figs.~\ref{Veff1} and \ref{Veff2} we deduce that the
size of the resulting shadow is
 sensitive on the \ae{}ther parameters $c_{13}$ and
$c_{14}$, and hence the image of M87* is able to impose   bounds on them.

In light of the report released by EHT collaboration for  the shadow of
M87* in
\cite{Akiyama:2019cqa,Akiyama:2019eap}, for the angular size of the shadow, the
mass and  the
distance to M87*   one respectively has the values
\begin{eqnarray}
 &&\delta = (42 \pm 3)\,\mu{ \text{arcsec}},\\
 &&M = (6.5 \pm 0.9) \times 10^9\,M_{\odot},\\
  &&D = 16.8^{+0.8}_{-0.7}\,{\text{ Mpc}},
\end{eqnarray}
where $M_{\odot}$ is the Sun mass.
One can
 merge this information by   introducing the single number $d_{M87*}$, which
quantifies the   size of M87*'s
shadow in unit   mass, defined as \cite{Bambi:2019tjh}
\begin{eqnarray}
d_{M87*} \equiv \frac{D\delta}{M} \approx 11.0 \pm 1.5\,.
\label{size}
\end{eqnarray}
This combination can be used in order to
  confront with the theoretically predicted shadows (e.g. the above number is
    in    agreement  within $1\sigma$-error
with what we theoretically expect for the Schwarzschild BH, namely $d_{
Schw}\simeq10.4$ \cite{Chandra}).

In order to confront Einstein-\AE ther solutions with the above observational
number, all we need is to calculate the  angular size $\delta_{I,II}$ for
each of the two types of solutions, as a function of the  Einstein-\AE ther
parameters. The angular size can be immediately extracted from the
   profile (\ref{fordelta}), as long as we know the radius of the unstable
photon sphere $r_{ps}$ (as a function of the  Einstein-\AE ther
parameters). Knowing $\delta_{I,II}$ can then easily lead to the predicted
diameter per  unit   mass
$d_{I,II}$ for each of the solutions.

\begin{figure}[ht]
	\includegraphics[width=8cm,height=6cm]{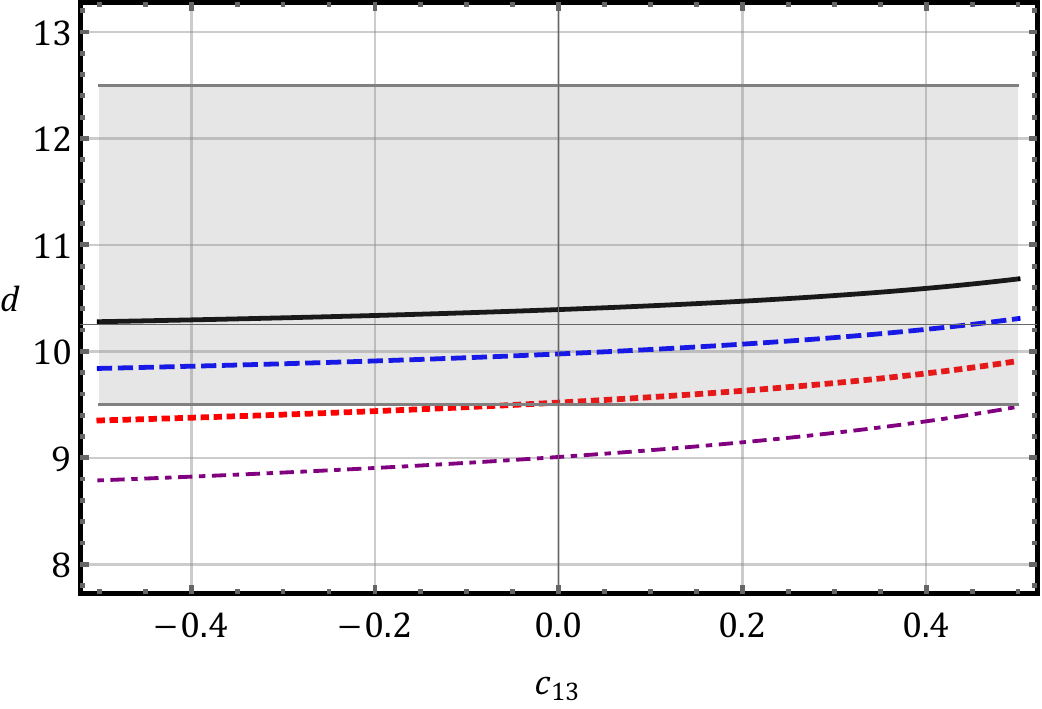}~~~~~
	\includegraphics[width=8cm,height=6cm]{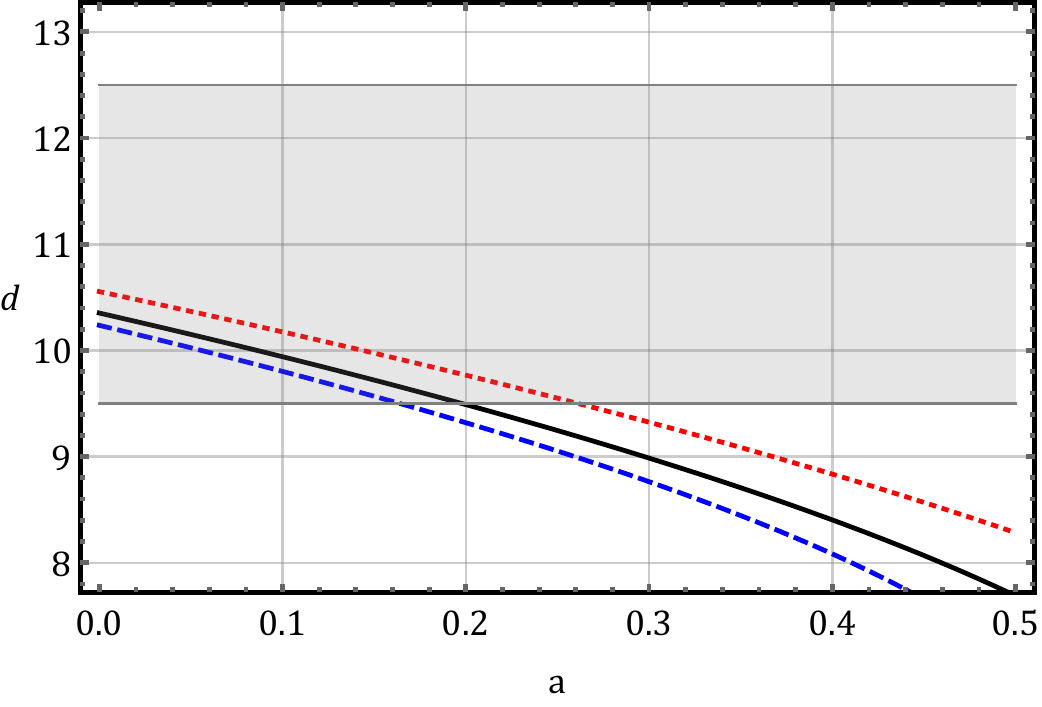}
	\caption{{\it{
		{\bf {Left graph:}}
The predicted diameter per  unit mass $d_I$ for the Einstein-\AE
ther type I black hole solution (\ref{slow1}) with (\ref{Sol1a}),
 as a function of the \ae{}ther parameter
$c_{13}$,  for several values of the rotational parameter: $a=0$ (black -
solid), $a=0.1$ (blue - dashed), $a=0.2$ (red - dotted), $a=0.3$ (purple -
dashed-dotted).
\textbf{Right graph:}
 $d_I$  as a function of the rotational parameter $a$, for  $c_{13}=-0.5$ (blue - dashed), $c_{13}=0$ (black- solid), $c_{13}=0.5$ (red- dotted).
In both graphs the shaded area mark the observationally determined
  diameter per unit mass of M87*'s shadow, namely  $d_{M87*}$, within
$1\sigma$-error.  }}}
\label{Scan1}
\end{figure}

\begin{figure}[ht]
 	\begin{center}
 \includegraphics[width=8cm,height=6cm]{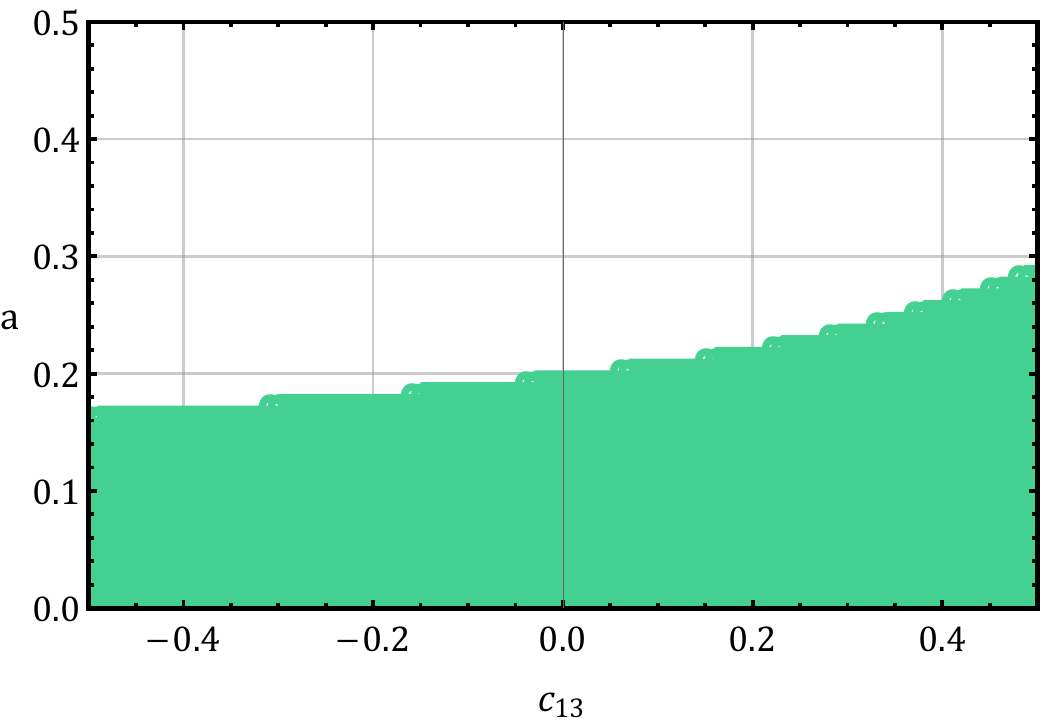}~~~~
 	\caption{{\it{The allowed parameter region ( green) in the 
 $c_{13}-a$ plane, for fixed BH mass ($M=6\times10^9 M_{\bigodot}$), for the
Einstein-\AE
ther type I black hole solution (\ref{slow1}) with (\ref{Sol1a}), that leads to
	diameter per  unit mass $d_I$  in agreement with the observationally
determined one   $d_{M87*}$  within $1\sigma$-error.}
}}
\label{Scan2}
\end{center}
\end{figure}

 Let us begin our analysis with Einstein-\AE ther  type I BH solution. 
Note that in agreement with EHT, from now on we 
fix the 
value $\theta_0=17$ for the inclination angle.
 In
 Fig.~\ref{Scan1} we depict the diameter of the predicted
 shadow $d_I$ as a function of $ c_{13}$ and $a$, on top of the observed one
 from
 M87* within $1\sigma$. As we observe,  \AE ther type I BH solution is
 able to quantitatively describe the shadow size of M87*, provided that the
 dimensionless  spin parameter  $a$ is constrained to specific values, dependent on value of $c_{13}$.
 Note that since our whole analysis has been restricted to the
 slow rotation case,   the extracted upper bounds on
 $a$ are  perfectly consistent and justify our approximation.
 The curves in Fig. \ref{Scan1}  imply that within these ranges of $c_{13}$ it is possible
 to distinguish a naked singularity
 ($c_{13}<0$) from BH ($c_{13}>0$).  Furthermore, in order to present the bounds
 on spin parameter  $a$ in a more transparent way, we perform a full scan of the two-dimensional
 parameter space and in Fig. \ref{Scan2}  we depict  the region which corresponds
 to a $d_I$ in agreement with the observed value $d_{M87*}$. As it is clear 
from 
it as well as from the right panel in Fig. \ref{Scan1} , by moving from 
negative to positive values of \ae{}ther parameter $c_{13}$, the upper bound on the 
$a$ grows. This may be useful in distinguishing between BH and naked 
singularity.  

\begin{figure}[ht]
	\begin{center}
		\includegraphics[width=8cm,height=6cm]{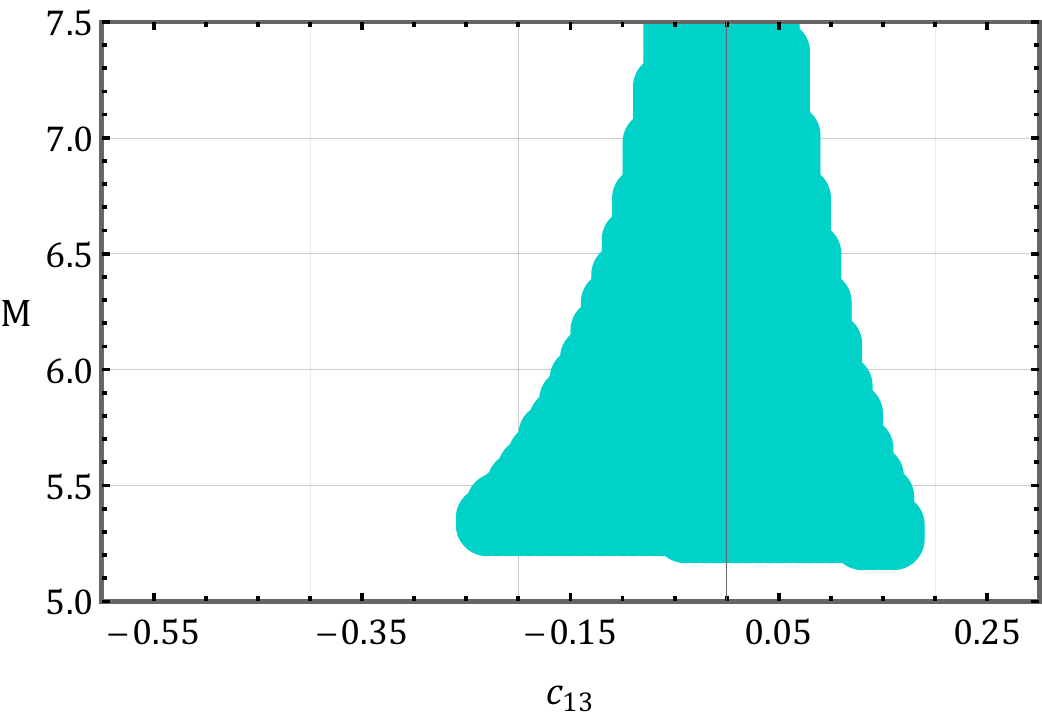}~~~
		\includegraphics[width=8cm,height=6cm]{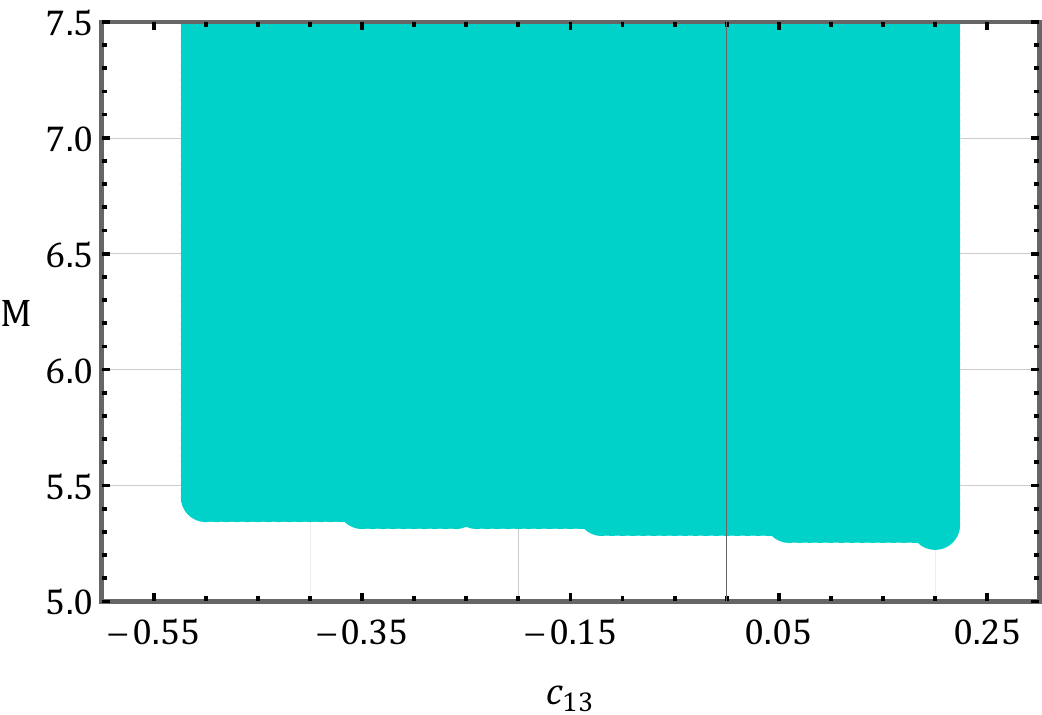}
		\caption{{\it{ The allowed parameter region (green) in 
the  $c_{13}-M(10^9 M_{\bigodot})$ plane, for fixed rotation parameter values  
$a=0$ (left graph) and $a=0.2$ (right graph),  for the
Einstein-\AE
ther type I black hole solution (\ref{slow1}) with (\ref{Sol1a}),  that leads to
					diameter $d_I$  in agreement with the observationally
					determined one   $d_{M87*}$  within $1\sigma$-error.}
		}}
		\label{Scan2m}
	\end{center}
\end{figure}

We mention  here that in order to create the allowed
parameter space  plots on the $c_{13}-a$ plane,  we have fixed the BH mass 
to $M=6\times 10^9 M_{\bigodot}$. 
Nevertheless, one can also perform a similar parameter scan for the case of 
non-fixed BH 
mass, namely within the two-dimensional parameter space $(c_{13}, M)$, for 
different fixed  values  of the rotation parameter $a$. In   Fig. \ref{Scan2m} 
we present such a graph. As we observe, for reasonable  $c_{13}$ values, the 
BH mass lies within the range we expect from EHT for M87*, i.e. $M= (6.5 \pm 
0.9) \times 10^9\,M_{\odot}$. Furthermore, Fig. \ref{Scan2m}   shows the 
role of rotation   on the allowed region of the  $c_{13}- M$ plane,
 namely  by 
increasing the rotation parameter   we are led to  larger region for the   
$c_{13}$ parameter  and still be consistent with  EHT data within 1-$\sigma$ 
error. 
In summary, Einstein-\AE ther  type I BH
solution is in agreement with M87* observation, and moreover we obtain a
restriction to low rotational values.


\begin{figure}[ht]
	\includegraphics[width=8cm,height=6cm]{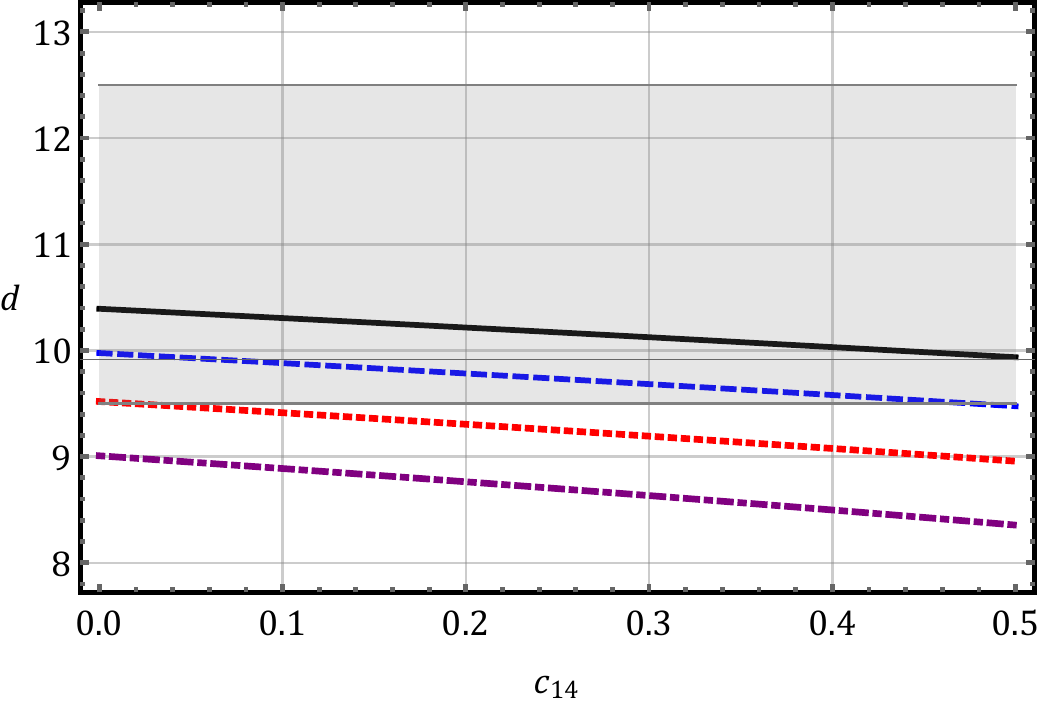}~~~~~
	\includegraphics[width=8cm,height=6cm]{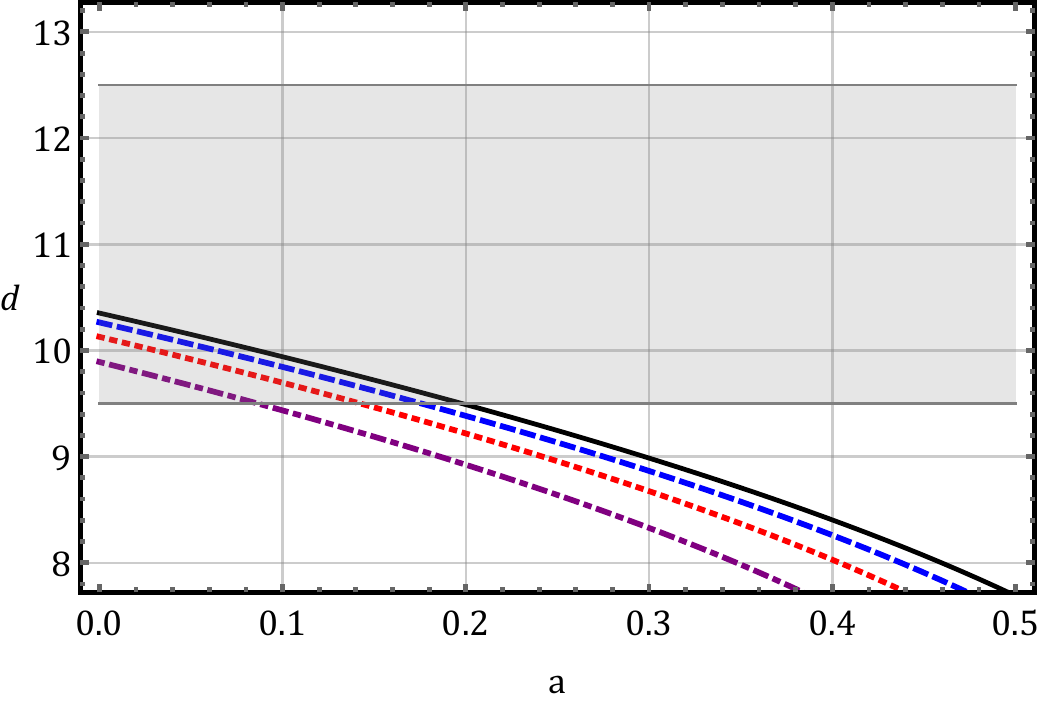}
	\caption{
	{\it{
		{\bf {Left graph:}}
The predicted diameter per  unit mass $d_{II}$ for the Einstein-\AE
ther type II black hole solution (\ref{slow1}) with (\ref{Sol2a}),
 as a function of the \ae{}ther parameter
$c_{14}$, for fixed $c_{13}=0$ and for several values of the rotational
parameter: $a=0$ (black-solid), $a=0.1$ (blue - dashed), $a=0.2$ (red - dotted), $a=0.3$
(purple -
dashed-dotted).
\textbf{Right graph:}
 $d_{II}$  as a function of the rotational parameter $a$,  for fixed
$c_{13}=0$ and for   $c_{14}=0$ (black-solid), $c_{14}=0.1$
(blue - solid), $c_{14}=0.25$ (red - dotted), $c_{14}=0.5$ (purple -
dashed-dotted). In both graphs the shaded area mark the observationally determined
  diameter per unit mass of M87*'s shadow, namely  $d_{M87*}$, within
$1\sigma$-error.}}
}	
\label{Scan3}
\end{figure}

\begin{figure}[ht]
	\includegraphics[width=8cm,height=6cm]{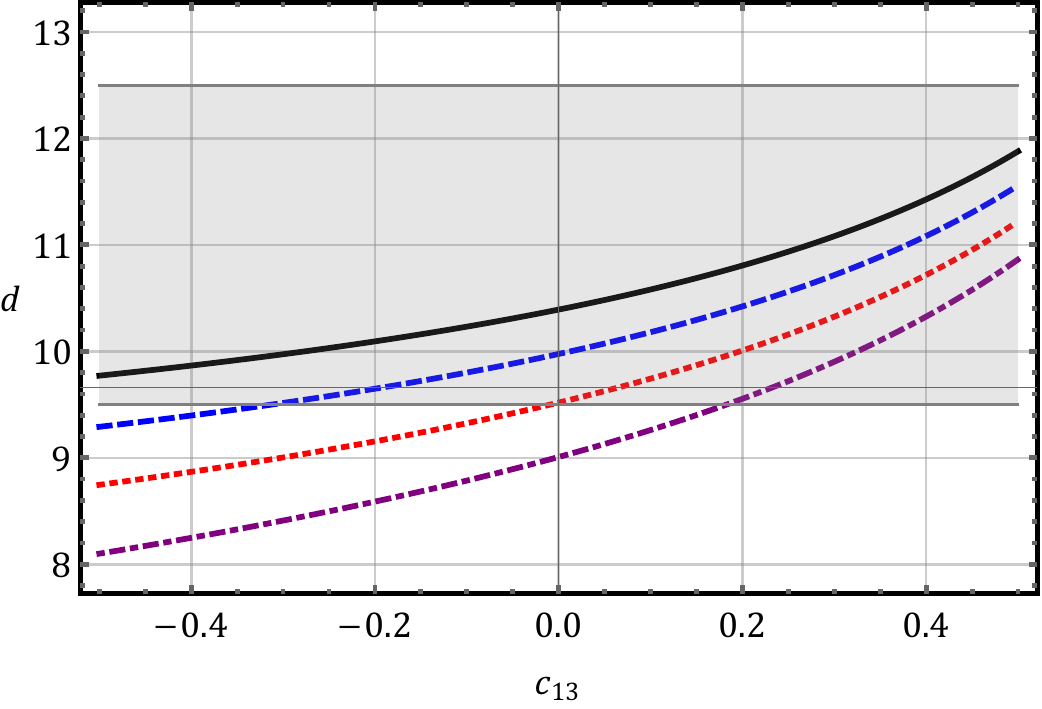}~~~
	\includegraphics[width=8cm,height=6cm]{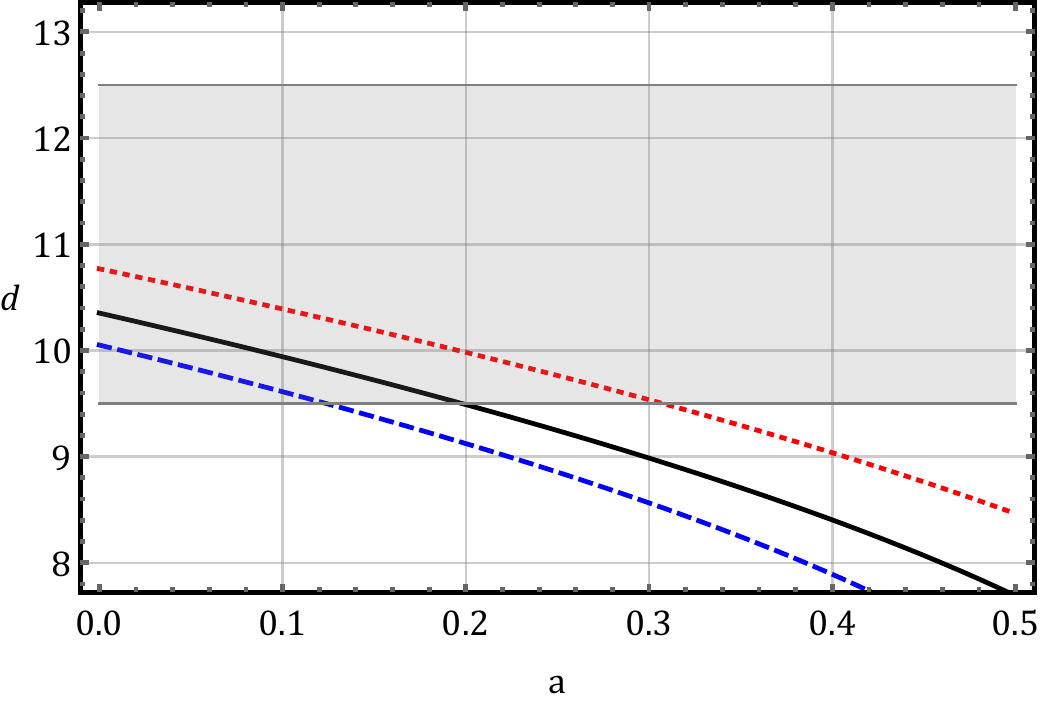}
	\caption{
	{\it{
		{\bf {Left graph:}}
The predicted diameter per  unit mass $d_{II}$ for the Einstein-\AE
ther type II black hole solution (\ref{slow1}) with (\ref{Sol2a}),
 as a function of the \ae{}ther parameter
$c_{13}$, for fixed $c_{14}=0$ and for several values of the rotational
parameter:  $a=0$ (black-solid), $a=0.1$ (blue - dashed), $a=0.2$ (red - dotted), $a=0.3$
(purple -
dashed-dotted).
\textbf{Right graph:}
 $d_{II}$  as a function of the rotational parameter $a$,  for fixed
$c_{14}=0$ and  for   $c_{13}=0$ (black-solid), $c_{13}=-0.2$
(blue - solid), $c_{13}=0.2$ (red - dotted).
In both graphs the shaded area mark the observationally determined
  diameter per unit mass of M87*'s shadow, namely  $d_{M87*}$, within
$1\sigma$-error.}}
}
\label{Scan4}
\end{figure}

 \begin{figure}[ht]
	\begin{center}
		\includegraphics[width=8cm,height=6cm]{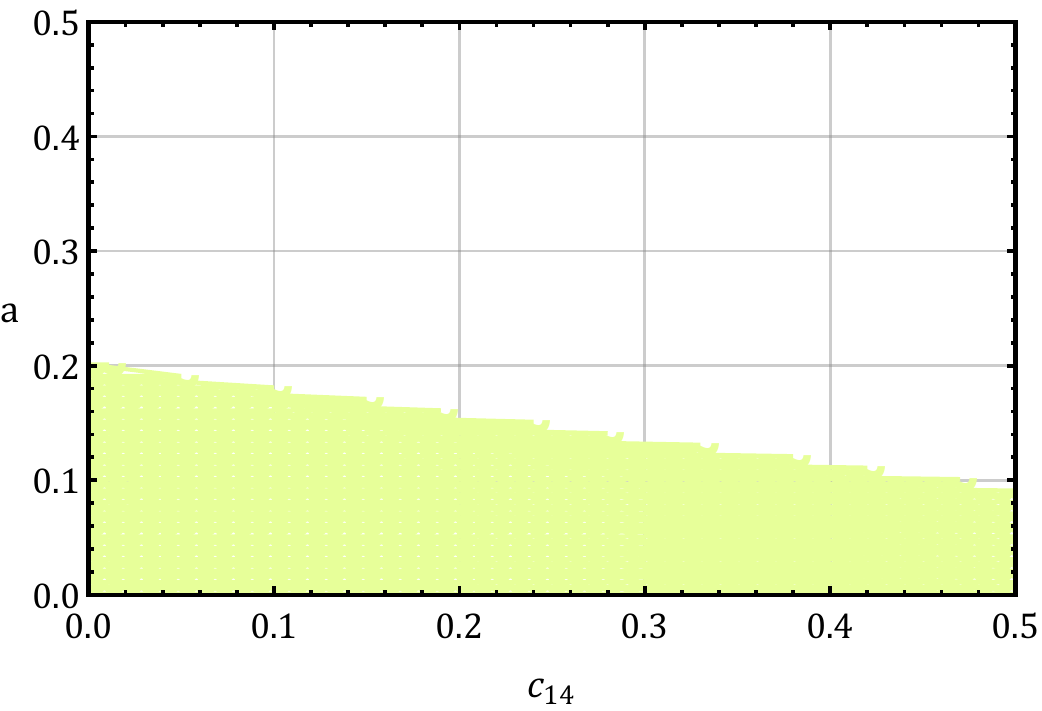}~~~
			\includegraphics[width=8cm,height=6cm]{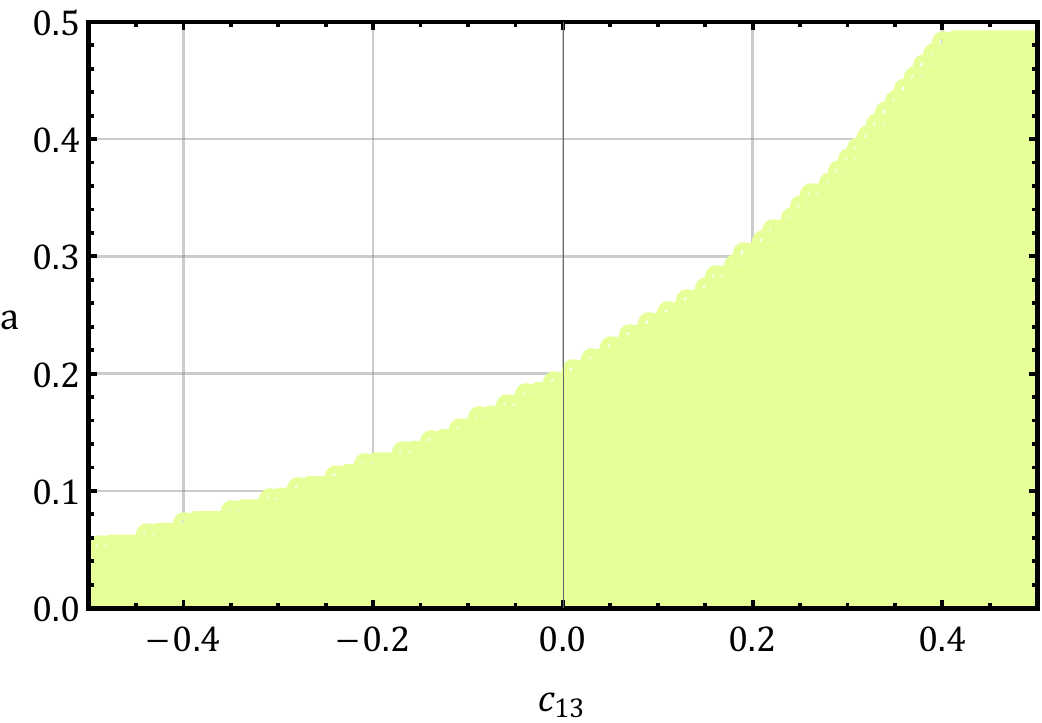}
		\caption{{\it{The allowed parameter regions (light 
green) in the  $c_{14}-a$ and  $c_{13}-a$ planes,  for fixed BH mass 
($M=6\times10^9 M_{\bigodot}$), for the
					Einstein-\AE
					ther type II black hole solution (\ref{slow1}) with (\ref{Sol2a}), that leads to
					diameter per  unit mass $d_{II}$  in agreement with the observationally
					determined one   $d_{M87*}$  within $1\sigma$-error.}}}
		\label{Scan5}
	\end{center}
\end{figure}

\begin{figure}[ht]
	\begin{center}
		\includegraphics[width=8cm,height=6cm]{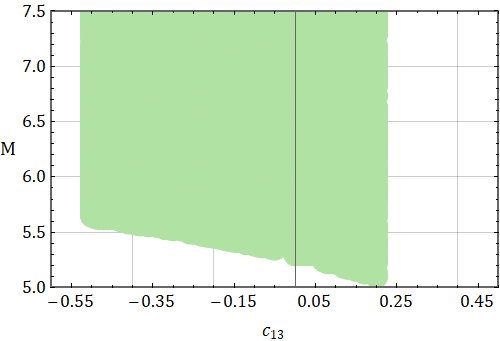}~~~
		\includegraphics[width=8cm,height=6cm]{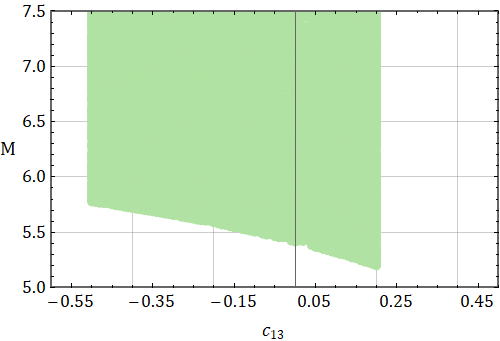}\\
		\includegraphics[width=8cm,height=6cm]{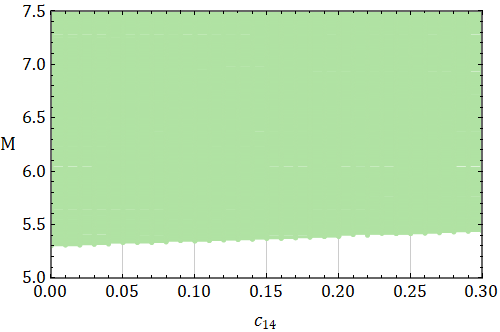}~~~
		\includegraphics[width=8cm,height=6cm]{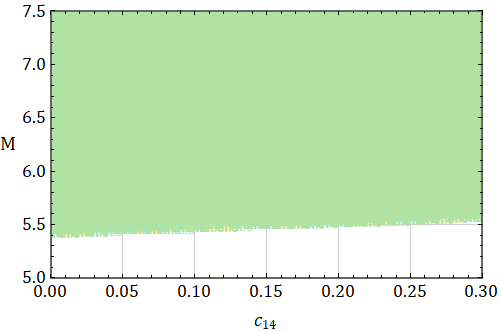}
		
		\caption{{\it{ The allowed parameter regions (green) in 
the  $c_{13}-M(10^9 M_{\bigodot})$ plane (upper graphs) and  $c_{14}-M(10^9 
M_{\bigodot})$ planes (lower graphs),
 for fixed rotation parameter values  $a=0$ 
(left graphs) and $a=0.2$ (right graphs),  for the
					Einstein-\AE
					ther type II black hole solution (\ref{slow1}) with (\ref{Sol2a}), that leads to
					diameter $d_{II}$  in agreement with the observationally
					determined one  $d_{M87*}$  within $1\sigma$-error.}}}
		\label{Scan6}
	\end{center}
\end{figure}

  We proceed to the investigation of the  Einstein-\AE ther type II BH solution,
which involves the  \ae{}ther parameters $c_{13}$
and $c_{14}$. In  Figs.~\ref{Scan3} and
\ref{Scan4}    we respectively draw the  diameter of the predicted
shadow as a function of $ c_{14}$,~$a$ and $c_{13}$,~$a$, on top of the observed one from
M87*
within $1\sigma$-error.
Similarly to the previous case, we deduce that we obtain agreement with the
observed behavior if  the involved  \ae{}ther parameters are restricted within
a given range, along with an upper bound on $a$. Concerning the \ae{}ther 
parameter $c_{14}$, as it increases then the upper bound on $a$ decreases, 
while for the parameter
$c_{13}$ this  $a$-behavior   happens by moving from positive values to 
negative value.
Additionally, the restriction of the rotation parameter $a$ to a given narrow 
window
can also be extracted by scanning the two-dimensional  spaces
$(c_{14},a)$ and $(c_{13},a)$, for fixed BH mass, as depicted in 
Fig.~\ref{Scan5}. Note that the  above statement on the change of
trend of \ae{}ther parameters and rotation parameter, can also be verified 
through   Fig.~\ref{Scan5}, too.
Finally, note that such small $a$ values verify that our slow
rotation approximation is indeed valid. 

Similarly to  the Einstein-\AE 	ther type I black hole 
solution, 
in the present type II solutions 
 we can also provide two-dimensional   scans on the planes $c_{13}- 
M$ and $c_{14}- M$, for fixed rotation parameters. 
As we observe in  
Fig.~\ref{Scan6}, unlike the type I solution case, in the present   Einstein-\AE
 	ther type II black hole solutions
  the rotation parameter   does not play a significant role 
in reducing the allowed parameter space. Moreover, 	the BH mass can be matched 
with what extracted by EHT for M87* within $1\sigma$-error.

We close this section by making some comments on the rotation behavior.
It is known that one might have
degeneracies of additional parameters
in  general relativity - magnetohydrodynamic (GR-MHD) simulations, which could
allow for
high spin values, namely  $a>0.5$
\cite{Akiyama:2019fyp}.
In particular, since    spin can be a strong energy source for
powering relativistic jets, one could have the case where
high  spin values  in specific
  GR-MHD models could produce
powerful jets.
The estimates of M87's jet power lie in the range $10^{42}$ to $10^{45} 
erg/s$ \cite{Reynolds:1996fh,deGasperin:2012id,Broderick:2015tda}, with values 
closer to   $10^{42}$ being favored, as it is observed form 
Table 3 of \cite{Akiyama:2019fyp}. Additionally, it is expected that the value 
of the real jet power could be higher than $10^{42} erg/s$  (for instance 
observations of Hubble Space Telescope (HST)-1 \cite{Stawarz:2006bk} and 
Low-Frequency
	Array (LOFAR) \cite{deGasperin:2012id} yield $\sim10^{44} erg/s$).  Hence, 
although   GR-MHD simulations  reveal that most of high spin models can produce 
sufficiently  powerful jets ($E > 10^{42} erg/s$)  and thus they can 
satisfy the lower limit of the power jets, they could 
still have no perfect agreement with the actual upper values. 
We mention here that  the jet formation mechanism in our scenario is  
the same with the case 
of general relativity, i.e we do not use 
  modified gravity as an explanation of the jet formation. A full
 MHD simulation in modified gravity  would be both necessary and interesting, 
nevertheless it  lies beyond the scope of the present work.
In general,   future high-fidelity interferometric imaging of
M87* and/or Sgr A*, are expected to clarify the role and value of spin.

For the moment it is adequate to resort to different probes of the BH rotation
rate, such as X-ray reflection spectroscopy.  In general, although observational
data indicate the existence of rotating
BHs,   there is no quantitative consensus  and the literature contains a
variety of
constraints
\cite{Reynolds:2020jwt,Reynolds:2013qqa}. For instance  it remains  an
open issue whether  the astrophysical BHs have
  slow, modest   or rapid rotational behavior.
Concerning supermassive BHs, the corresponding information is obtained through
the relativistic X-ray reflection spectroscopy, which is a
powerful technique applicable to measure robust
BH spins across the mass range, from the stellar-mass BHs in X-ray binaries to
the supermassive BHs in active galactic nuclei (AGN)
\cite{Reynolds:2020jwt,Reynolds:2013qqa}. By utilizing this   method
within recent structure formation models, one might face
a mass dependency for
the distribution of supermassive BH spins
\cite{Volonteri:2004cf,Sesana:2014bea,Zhang:2019ekx,Bustamante:2019sjz}.
However, an important point to note is that the
X-ray reflection technique should be  used only for the
standard accretion disk case, i.e. for moderate mass accretion rates in 
the range $\dot{M}\sim 0.01-0.3 \dot{M}_{Edd}$, with $\dot{M}_{Edd}$ the 
Eddington mass accretion rate \cite{Reynolds:2020jwt}. Although it is expected 
that this technique is applicable for BH X-ray binaries in their luminous hard 
state,    it cannot work for the low-luminosity type of AGNs 
like M87* and Sgr A*, since the corresponding accretion rate   is very low.

\section{Conclusions}\label{co}

In the present work we   investigated  Einstein-\AE ther  gravity  in light of
the  recent  Event Horizon Telescope (EHT) observations of the M87*.  In
particular, the EHT provided the first visual evidence indicating directly the
existence of a compact object  such as a supermassive black hole (BH) candidate
at the center of the  M87* galaxy. Since  the shape and size of the observed BH
shadow contains information of the geometry in the vicinity  of the BH,
one can   consider  the shadow as a potential probe to investigate the BH
structure within different   gravitational theories, due to the fact that in
modified gravity one obtain black holes which deviate from those of general
relativity. Hence,  allowing for a modified gravity as the underlying theory
enriches the calculation framework with  different size-rotation features as
well as with extra model parameters.

The Einstein-\AE ther (EA)  theory   is a generally covariant theory
of gravity, which violates the Lorentz invariance locally by possessing a
dynamical, unit-norm and timelike vector field, namely the \ae{}ther
field, which defines a preferred timelike direction at each
spacetime point. The black hole solutions in such a framework include
two classes of slow rotating spherically symmetric   solutions, depending on   the
involved mixed \ae{}ther parameters $c_{13}$ and $c_{14}$.

We first extracted the above  black hole solutions for EA gravity, and we
calculated the corresponding    effective potential
  $U_{eff}(r)$ for the photons,   the resulting event horizon
radius    $r_e$  and the  radius of the unstable photon sphere $r_{ps}$. Then
we
straightforwardly calculated the induced angular size $\delta$, which combined
with the  mass and  the distance can lead to a single prediction
that quantifies the black hole shadow size, namely the diameter per  unit mass
 $d$. Since  $d_{M87*}$ is
observationally known from the EHT Probe, we extracted the corresponding
parameter regions of instein-\AE ther theory in order to obtain consistency.

 Apart from the restriction of the \ae{}ther parameters $c_{13}$ 
and $c_{14}$ to specif ranges, we found some constraints on the dimensionless 
spin parameter $a$ in slow rotation limit, dependent on the values of \ae{}ther 
parameters,  which was verified by a full scan of the parameter space. 
Furthermore, our analysis   has indicated that in the Einstein-\AE ther (EA)  
theory (type I BH solution)   is possible to distinguish a naked singularity 
from BH.  
This is indeed a verification of the fact that in modified gravity one obtains
 different size-rotation features, depending on the extra model parameters, and
hence the  M87* observation can be in principle described in different ways
comparing to general relativity, especially in theories
 which possess various corrections on the  Kerr-metric solutions,
such is the case in Einstein-\AE ther gravity.

In summary, Einstein-\AE ther black hole solutions are in agreement with EHT
M87* observation,  and this may act as an advantage for Einstein-\AE ther
gravity.

\begin{acknowledgments}

M. Kh. would like to thank H. Firouzjahi for useful discussions.  E.N.S's work
is supported in part by the USTC
Fellowship for international professors.

\end{acknowledgments}


\providecommand{\href}[2]{#2}\begingroup\raggedright


\begin{thebibliography}{100}



	
\bibitem{Shakura:1972te}
N.~I.~Shakura and R.~A.~Sunyaev,
{\it{Black holes in binary systems. Observational appearance}},
Astron.\ Astrophys.\  {\bf 24}, 337 (1973).	
	
\bibitem{Abbott:2016blz}
B.~P.~Abbott {\it et al.} [LIGO Scientific and Virgo Collaborations],
{\it{Observation of Gravitational Waves from a Binary Black Hole Merger}},
Phys.\ Rev.\ Lett.\  {\bf 116}, no. 6, 061102 (2016)
[\href{http://xxx.lanl.gov/abs/1602.03837}
{{\tt arXiv:1602.03837}}].





\bibitem{Abbott:2016nmj}
B.~P.~Abbott {\it et al.} [LIGO Scientific and Virgo Collaborations],
{\it{GW151226: Observation of Gravitational Waves from a 22-Solar-Mass Binary
Black Hole Coalescence}},
Phys.\ Rev.\ Lett.\  {\bf 116}, no. 24, 241103 (2016)
[\href{http://xxx.lanl.gov/abs/1606.04855}
{{\tt arXiv:1606.04855}}].




\bibitem{Abbott:2017vtc}
B.~P.~Abbott {\it et al.} [LIGO Scientific and VIRGO Collaborations],
{\it{GW170104: Observation of a 50-Solar-Mass Binary Black Hole Coalescence at
Redshift 0.2}},
Phys.\ Rev.\ Lett.\  {\bf 118}, no. 22, 221101 (2017)
Erratum: [Phys.\ Rev.\ Lett.\  {\bf 121}, no. 12, 129901 (2018)]
[\href{http://xxx.lanl.gov/abs/1706.01812}
{{\tt arXiv:1706.01812}}].



\bibitem{Abbott:2017oio}
B.~P.~Abbott {\it et al.} [LIGO Scientific and Virgo Collaborations],
{\it{GW170814: A Three-Detector Observation of Gravitational Waves from a
Binary Black Hole Coalescence}},
Phys.\ Rev.\ Lett.\  {\bf 119}, no. 14, 141101 (2017)
[\href{http://xxx.lanl.gov/abs/1709.09660}
{{\tt arXiv:1709.09660}}].



\bibitem{Abbott:2017gyy}
B.~. P.~.Abbott {\it et al.} [LIGO Scientific and Virgo Collaborations],
{\it{GW170608: Observation of a 19-solar-mass Binary Black Hole Coalescence}},
Astrophys.\ J.\ Lett.\  {\bf 851}, L35 (2017)
[\href{http://xxx.lanl.gov/abs/1711.05578}
{{\tt arXiv:1711.05578}}].



\bibitem{TheLIGOScientific:2017qsa}
B.~P.~Abbott {\it et al.} [LIGO Scientific and Virgo Collaborations],
{\it{GW170817: Observation of Gravitational Waves from a Binary Neutron Star
Inspiral}},
Phys.\ Rev.\ Lett.\  {\bf 119}, no. 16, 161101 (2017)
[\href{http://xxx.lanl.gov/abs/1710.05832}
{{\tt arXiv:1710.05832}}].




\bibitem{Monitor:2017mdv}
B.~P.~Abbott {\it et al.} [LIGO Scientific and Virgo and Fermi-GBM and INTEGRAL
Collaborations],
{\it{Gravitational Waves and Gamma-rays from a Binary Neutron Star Merger:
GW170817 and GRB 170817A}},
Astrophys.\ J.\ Lett.\  {\bf 848}, no. 2, L13 (2017)
[\href{http://xxx.lanl.gov/abs/1710.05834}
{{\tt arXiv:1710.05834}}].



\bibitem{Akiyama:2019cqa}
K.~Akiyama {\it et al.} [Event Horizon Telescope Collaboration],
{\it{First M87 Event Horizon Telescope Results. I. The Shadow of the
Supermassive Black Hole}},
Astrophys.\ J.\ Lett.\  {\bf 875}, L1 (2019)
[\href{http://xxx.lanl.gov/abs/1906.11238}
{{\tt arXiv:1906.11238}}].


\bibitem{Akiyama:2019brx}
K.~Akiyama {\it et al.} [Event Horizon Telescope Collaboration],
{\it{First M87 Event Horizon Telescope Results. II. Array and Instrumentation}},
Astrophys.\ J.\ Lett.\  {\bf 875}, no. 1, L2 (2019)
[\href{http://xxx.lanl.gov/abs/1906.11239}
{{\tt arXiv:1906.11239}}].



\bibitem{Akiyama:2019sww}
K.~Akiyama {\it et al.} [Event Horizon Telescope Collaboration],
{\it{First M87 Event Horizon Telescope Results. III. Data Processing and
Calibration}},
Astrophys.\ J.\ Lett.\  {\bf 875}, no. 1, L3 (2019)
[\href{http://xxx.lanl.gov/abs/1906.11240}
{{\tt arXiv:1906.11240}}].



\bibitem{Akiyama:2019bqs}
K.~Akiyama {\it et al.} [Event Horizon Telescope Collaboration],
{\it{First M87 Event Horizon Telescope Results. IV. Imaging the Central
Supermassive Black Hole}},
Astrophys.\ J.\ Lett.\  {\bf 875}, no. 1, L4 (2019)
[\href{http://xxx.lanl.gov/abs/1906.11241}
{{\tt arXiv:1906.11241}}].




\bibitem{Akiyama:2019fyp}
K.~Akiyama {\it et al.} [Event Horizon Telescope Collaboration],
{\it{First M87 Event Horizon Telescope Results. V. Physical Origin of the
Asymmetric Ring}},
Astrophys.\ J.\ Lett.\  {\bf 875}, no. 1, L5 (2019)
[\href{http://xxx.lanl.gov/abs/1906.11242}
{{\tt arXiv:1906.11242}}].


\bibitem{Akiyama:2019eap}
K.~Akiyama {\it et al.} [Event Horizon Telescope Collaboration],
{\it{First M87 Event Horizon Telescope Results. VI. The Shadow and Mass of the
Central Black Hole}},
Astrophys.\ J.\ Lett.\  {\bf 875}, no. 1, L6 (2019)
[\href{http://xxx.lanl.gov/abs/1906.11243}
{{\tt arXiv:1906.11243}}].



\bibitem{Ohgami:2015nra}
T.~Ohgami and N.~Sakai,
{\it{Wormhole shadows}},
Phys.\ Rev.\ D {\bf 91}, no. 12, 124020 (2015)
[\href{http://xxx.lanl.gov/abs/1704.07065}
{{\tt arXiv:1704.07065}}].



\bibitem{Shaikh:2018kfv}
R.~Shaikh,
{\it{Shadows of rotating wormholes}},
Phys.\ Rev.\ D {\bf 98}, no. 2, 024044 (2018)
[\href{http://xxx.lanl.gov/abs/1803.11422}
{{\tt arXiv:1803.11422}}].



\bibitem{Virbhadra:2002ju} 
K.~S.~Virbhadra and G.~F.~R.~Ellis,
{\it{Gravitational lensing by naked singularities}},
Phys.\ Rev.\ D {\bf 65}, 103004 (2002).


\bibitem{Virbhadra:2007kw} 
K.~S.~Virbhadra and C.~R.~Keeton,
{\it{Time delay and magnification centroid due to gravitational lensing by black holes and naked singularities}},
Phys.\ Rev.\ D {\bf 77}, 124014 (2008)
[\href{http://xxx.lanl.gov/abs/0710.2333}
{{\tt arXiv:0710.2333}}].




\bibitem{Ortiz:2015rma}
N.~Ortiz, O.~Sarbach and T.~Zannias,
{\it{Shadow of a naked singularity}},
Phys.\ Rev.\ D {\bf 92}, no. 4, 044035 (2015)
[\href{http://xxx.lanl.gov/abs/1505.07017}
{{\tt arXiv:1505.07017}}].



\bibitem{Shaikh:2018lcc}
R.~Shaikh, P.~Kocherlakota, R.~Narayan and P.~S.~Joshi,
{\it{Shadows of spherically symmetric black holes and naked singularities}},
Mon.\ Not.\ Roy.\ Astron.\ Soc.\  {\bf 482}, no. 1, 52 (2019)
[\href{http://xxx.lanl.gov/abs/1802.08060}
{{\tt arXiv:1802.08060}}].



\bibitem{Joshi:2020tlq}
A.~B.~Joshi, D.~Dey, P.~S.~Joshi and P.~Bambhaniya,
{\it{Shadow of a Naked Singularity without Photon Sphere}},
Phys.\ Rev.\ D {\bf 102}, no. 2, 024022 (2020)
[\href{http://xxx.lanl.gov/abs/2004.06525}
{{\tt arXiv:2004.06525}}].




\bibitem{Broderick:2009ph}
  A.~E.~Broderick, A.~Loeb and R.~Narayan,
  {\it{The Event Horizon of Sagittarius A*}},
  Astrophys.\ J.\  {\bf 701}, 1357 (2009)
  [\href{http://xxx.lanl.gov/abs/0903.1105}
{{\tt arXiv:0903.1105}}].


\bibitem{Bambi:2012bh}
  C.~Bambi,
  {\it{A note on the observational evidence for the existence of event horizons
in astrophysical black hole candidates}},
  Scientific World Journal {\bf 2013}, 204315 (2013)
    [\href{http://xxx.lanl.gov/abs/1205.4640}
{{\tt arXiv:1205.4640}}].



\bibitem{Cunha:2018acu}
  P.~V.~P.~Cunha and C.~A.~R.~Herdeiro,
{\it{Shadows and strong gravitational lensing: a brief review}},
  Gen.\ Rel.\ Grav.\  {\bf 50}, no. 4, 42 (2018)
      [\href{http://xxx.lanl.gov/abs/1801.00860}
{{\tt arXiv:1801.00860}}].




\bibitem{Chandra}
S. Chandrasekhar, {\it{The mathematical theory of black holes}},
Oxford classic texts in the physical
sciences, Oxford Univ. Press, Oxford (2002).

\bibitem{Synge:1966okc}
J.~L.~Synge,
{\it{The Escape of Photons from Gravitationally Intense Stars}},
Mon.\ Not.\ Roy.\ Astron.\ Soc.\  {\bf 131}, no. 3, 463 (1966).

\bibitem{Luminet:1979nyg}
J.-P.~Luminet,
{\it{Image of a spherical black hole with thin accretion disk}},
Astron.\ Astrophys.\  {\bf 75}, 228 (1979).

\bibitem{Bardeen:1973}
J. M. Bardeen, {\it{Timelike and null geodesics of the Kerr metric}}, Gordon
Breach, Science Publishers, New York (1973).

\bibitem{Narayan:2019imo}
  R.~Narayan, M.~D.~Johnson and C.~F.~Gammie,
{\it{The Shadow of a Spherically Accreting Black Hole}},
  Astrophys.\ J.\ Lett.\  {\bf 885}, no. 2, L33 (2019)
        [\href{http://xxx.lanl.gov/abs/1910.02957}
{{\tt arXiv:1910.02957}}].



\bibitem{Falcke:1999pj}
  H.~Falcke, F.~Melia and E.~Agol,
{\it{Viewing the shadow of the black hole at the galactic center}},
  Astrophys.\ J.\ Lett.\  {\bf 528}, L13 (2000)
          [\href{http://xxx.lanl.gov/abs/astro-ph/9912263}
{{\tt arXiv:astro-ph/9912263}}].


\bibitem{Younsi:2016azx} 
Z.~Younsi, A.~Zhidenko, L.~Rezzolla, R.~Konoplya and Y.~Mizuno,
{\it{New method for shadow calculations: Application to parametrized axisymmetric black holes}},
Phys.\ Rev.\ D {\bf 94}, no. 8, 084025 (2016)
[\href{http://xxx.lanl.gov/abs/1607.05767}
{{\tt arXiv:1607.05767}}].


\bibitem{Mizuno:2018lxz} 
Y.~Mizuno, Z.~Younsi, C.~M.~Fromm, O.~Porth, M.~De Laurentis, H.~Olivares, 
H.~Falcke, M.~Kramer and L.~Rezzolla,
{\it{The Current Ability to Test Theories of Gravity with Black Hole Shadows}},
Nature Astron.\  {\bf 2}, no. 7, 585 (2018)
 [\href{http://xxx.lanl.gov/abs/1804.05812}
{{\tt arXiv:1804.05812}}].

 
\bibitem{DeFalco:2021klh} 
V.~De Falco, E.~Battista, S.~Capozziello and M.~De Laurentis,
{\it {Testing wormhole solutions in extended gravity through the Poynting-Robertson effect}},
Phys.\ Rev.\ D {\bf 103}, no. 4, 044007 (2021)
 [\href{http://xxx.lanl.gov/abs/2101.04960}
{{ \tt arXiv:2101.04960}}].

  \bibitem{Shaikh:2019fpu}
  R.~Shaikh,
  {\it{Black hole shadow in a general rotating spacetime obtained through
Newman-Janis algorithm}},
  Phys.\ Rev.\ D {\bf 100}, no. 2, 024028 (2019)
            [\href{http://xxx.lanl.gov/abs/1904.08322}
{{\tt arXiv:1904.08322}}].


\bibitem{Wei:2019pjf}
  S.~W.~Wei, Y.~C.~Zou, Y.~X.~Liu and R.~B.~Mann,
{\it{Curvature radius and Kerr black hole shadow}},
  JCAP {\bf 1908}, 030 (2019)
              [\href{http://xxx.lanl.gov/abs/1904.07710}
{{\tt arXiv:1904.07710}}].




\bibitem{Moffat:2019uxp}
  J.~W.~Moffat and V.~T.~Toth,
{\it{Masses and shadows of the black holes Sagittarius A* and M87* in modified
gravity}},
  Phys.\ Rev.\ D {\bf 101}, no. 2, 024014 (2020)
                [\href{http://xxx.lanl.gov/abs/1904.04142}
{{\tt arXiv:1904.04142}}].




\bibitem{Firouzjaee:2019aij}
  J.~T.~Firouzjaee and A.~Allahyari,
{\it{Black hole shadow with a cosmological constant for cosmological
observers}},
  Eur.\ Phys.\ J.\ C {\bf 79}, no. 11, 930 (2019)
                  [\href{http://xxx.lanl.gov/abs/1905.07378}
{{\tt arXiv:1905.07378}}].


\bibitem{Banerjee:2019cjk}
  I.~Banerjee, B.~Mandal and S.~SenGupta,
  {\it{Does black hole continuum spectrum signal $f(R)$ gravity in higher
dimensions?}},
  Phys.\ Rev.\ D {\bf 101}, no. 2, 024013 (2020)
                    [\href{http://xxx.lanl.gov/abs/1905.12820}
{{\tt arXiv:1905.12820}}].



 \bibitem{Long:2019nox}
  F.~Long, J.~Wang, S.~Chen and J.~Jing,
 {\it{Shadow of a rotating squashed Kaluza-Klein black hole}},
  JHEP {\bf 1910}, 269 (2019)
                      [\href{http://xxx.lanl.gov/abs/1906.04456}
{{\tt arXiv:1906.04456}}].


 \bibitem{Zhu:2019ura}
  T.~Zhu, Q.~Wu, M.~Jamil and K.~Jusufi,
  {\it{Shadows and deflection angle of charged and slowly rotating black holes
in Einstein-Æther theory}},
  Phys.\ Rev.\ D {\bf 100}, no. 4, 044055 (2019)
                       [\href{http://xxx.lanl.gov/abs/1906.05673}
{{\tt arXiv:1906.05673}}].




\bibitem{Konoplya:2019goy}
  R.~A.~Konoplya and A.~Zhidenko,
{\it{Analytical representation for metrics of scalarized Einstein-Maxwell black
holes and their shadows}},
  Phys.\ Rev.\ D {\bf 100}, no. 4, 044015 (2019)
                         [\href{http://xxx.lanl.gov/abs/1907.05551}
{{\tt arXiv:1907.05551}}].




\bibitem{Contreras:2019cmf}
  E.~Contreras, Á.~Rincón, G.~Panotopoulos, P.~Bargueño and B.~Koch,
{\it{Black hole shadow of a rotating scale--dependent black hole}},
  Phys.\ Rev.\ D {\bf 101}, no. 6, 064053 (2020)
                           [\href{http://xxx.lanl.gov/abs/1906.06990}
{{\tt arXiv:1906.06990}}].



\bibitem{Li:2020drn}
  P.~C.~Li, M.~Guo and B.~Chen,
{\it{Shadow of a Spinning Black Hole in an Expanding Universe}},
  Phys.\ Rev.\ D {\bf 101}, no. 8, 084041 (2020)
      [\href{http://xxx.lanl.gov/abs/2001.04231}
{{\tt arXiv:2001.04231}}].


\bibitem{Kumar:2020pol}
  R.~Kumar, S.~G.~Ghosh and A.~Wang,
{\it{Gravitational deflection of light and shadow cast by rotating Kalb-Ramond
black holes}},
  Phys.\ Rev.\ D {\bf 101}, no. 10, 104001 (2020)
                               [\href{http://xxx.lanl.gov/abs/2001.00460}
{{\tt arXiv:2001.00460}}].




\bibitem{Pantig:2020uhp}
  R.~C.~Pantig and E.~T.~Rodulfo,
  {\it{Rotating dirty black hole and its shadow}},
  Chin.\ J.\ Phys.\  {\bf 68}, 236 (2020)
                                 [\href{http://xxx.lanl.gov/abs/2003.06829}
{{\tt arXiv:2003.06829}}].



\bibitem{Xavier:2020egv}
  S.~V.~M.~C.~B.~Xavier, P.~V.~P.~Cunha, L.~C.~B.~Crispino and
C.~A.~R.~Herdeiro,
{\it{Shadows of charged rotating black holes: Kerr–Newman versus Kerr–Sen}},
  Int.\ J.\ Mod.\ Phys.\ D {\bf 29}, no. 11, 2041005 (2020)
                                   [\href{http://xxx.lanl.gov/abs/2003.14349}
{{\tt arXiv:2003.14349}}].






 \bibitem{Guo:2020zmf}
  M.~Guo and P.~C.~Li,
{\it{Innermost stable circular orbit and shadow of the $4D$
Einstein–Gauss–Bonnet black hole}},
  Eur.\ Phys.\ J.\ C {\bf 80}, no. 6, 588 (2020)
                                   [\href{http://xxx.lanl.gov/abs/2003.02523}
{{\tt arXiv:2003.02523}}].



\bibitem{Roy:2020dyy}
  R.~Roy and S.~Chakrabarti,
{\it{Study on black hole shadows in asymptotically de Sitter spacetimes}},
  Phys.\ Rev.\ D {\bf 102}, no. 2, 024059 (2020)
   [\href{http://xxx.lanl.gov/abs/2003.14107}
{{\tt arXiv:2003.14107}}].



 \bibitem{Jin:2020emq}
  X.~H.~Jin, Y.~X.~Gao and D.~J.~Liu,
 {\it{Strong gravitational lensing of a 4-dimensional Einstein–Gauss–Bonnet
black hole in homogeneous plasma}},
  Int.\ J.\ Mod.\ Phys.\ D {\bf 29}, no. 09, 2050065 (2020)
     [\href{http://xxx.lanl.gov/abs/2004.02261}
{{\tt arXiv:2004.02261}}].



\bibitem{Islam:2020xmy}
  S.~U.~Islam, R.~Kumar and S.~G.~Ghosh,
{\it{Gravitational lensing by black holes in the $4D$ Einstein-Gauss-Bonnet
gravity}},
  JCAP {\bf 2009}, 030 (2020)
       [\href{http://xxx.lanl.gov/abs/2004.01038}
{{\tt arXiv:2004.01038}}].

\bibitem{Chen:2020aix}
C.~Y.~Chen,
{\it{Rotating black holes without $\mathbb{Z}_2$ symmetry and their shadow
images}},
JCAP \textbf{05}, 040 (2020)
      [\href{http://xxx.lanl.gov/abs/2004.01440}
{{\tt arXiv:2004.01440}}].




\bibitem{Zeng:2020vsj}
  X.~X.~Zeng and H.~Q.~Zhang,
{\it{Influence of quintessence dark energy on the shadow of black hole}},
Eur. Phys. J. C \textbf{80}, no.11, 1058,
       [\href{http://xxx.lanl.gov/abs/2007.06333}
{{\tt arXiv:2007.06333}}].



\bibitem{Konoplya:2020xam}
  R.~A.~Konoplya, J.~Schee and D.~Ovchinnikov,
{\it{Shadow of the magnetically and tidally deformed black hole}},
       [\href{http://xxx.lanl.gov/abs/2008.04118}
{{\tt arXiv:2008.04118}}].



\bibitem{Belhaj:2020mlv}
  A.~Belhaj, M.~Benali, A.~E.~Balali, W.~E.~Hadri, H.~El Moumni and
E.~Torrente-Lujan,
{\it{Black Hole Shadows in M-theory Scenarios}},
       [\href{http://xxx.lanl.gov/abs/2008.09908}
{{\tt arXiv:2008.09908}}].




\bibitem{Long:2020wqj}
  F.~Long, S.~Chen, M.~Wang and J.~Jing,
{\it{Shadow of a disformal Kerr black hole in quadratic DHOST theories}},
       [\href{http://xxx.lanl.gov/abs/2009.07508}
{{\tt arXiv:2009.07508}}].




\bibitem{Jusufi:2020zln}
  K.~Jusufi and Saurabh,
{\it{Black Hole Shadows in Verlinde's Emergent Gravity}},
       [\href{http://xxx.lanl.gov/abs/2010.15870}
{{\tt arXiv:2010.15870}}].



\bibitem{Contreras:2020kgy}
  E.~Contreras, Á.~Rincón, G.~Panotopoulos and P.~Bargueño,
{\it{Geodesic analysis and black hole shadows on a general non--extremal
rotating black hole in five--dimensional gauged supergravity}},
     [\href{http://xxx.lanl.gov/abs/2010.03734}
{{\tt arXiv:2010.03734}}].

\bibitem{Shao:2020weq}
W.~H.~Shao, C.~Y.~Chen and P.~Chen,
{\it{Generating Rotating Spacetime in Ricci-Based Gravity: Naked Singularity as
a Black Hole Mimicker}},
     [\href{http://xxx.lanl.gov/abs/2011.07763}
{{\tt arXiv:2011.07763}}].


\bibitem{Ghosh:2020spb}
  S.~G.~Ghosh, R.~Kumar and S.~U.~Islam,
{\it{Parameters estimation and strong gravitational lensing of nonsingular
Kerr-Sen black holes}},
     [\href{http://xxx.lanl.gov/abs/2011.08023}
{{\tt arXiv:2011.08023}}].




\bibitem{Davoudiasl:2019nlo}
  H.~Davoudiasl and P.~B.~Denton,
{\it{Ultralight Boson Dark Matter and Event Horizon Telescope Observations of
M87*}},
  Phys.\ Rev.\ Lett.\  {\bf 123}, no. 2, 021102 (2019)
         [\href{http://xxx.lanl.gov/abs/1904.09242}
{{\tt arXiv:1904.09242}}].



\bibitem{Bar:2019pnz}
  N.~Bar, K.~Blum, T.~Lacroix and P.~Panci,
{\it{Looking for ultralight dark matter near supermassive black holes}},
  JCAP {\bf 1907}, 045 (2019)
           [\href{http://xxx.lanl.gov/abs/1905.11745}
{{\tt arXiv:1905.11745}}].




\bibitem{Jusufi:2019nrn}
  K.~Jusufi, M.~Jamil, P.~Salucci, T.~Zhu and S.~Haroon,
{\it{Black Hole Surrounded by a Dark Matter Halo in the M87 Galactic Center and
its Identification with Shadow Images}},
  Phys.\ Rev.\ D {\bf 100}, no. 4, 044012 (2019)
             [\href{http://xxx.lanl.gov/abs/1905.11803}
{{\tt arXiv:1905.11803}}].


  \bibitem{Konoplya:2019sns}
  R.~A.~Konoplya,
 {\it{Shadow of a black hole surrounded by dark matter}},
  Phys.\ Lett.\ B {\bf 795}, 1 (2019)
               [\href{http://xxx.lanl.gov/abs/1905.00064}
{{\tt arXiv:1905.00064}}].




\bibitem{Narang:2020bgo}
  A.~Narang, S.~Mohanty and A.~Kumar,
{\it{Test of Kerr-Sen metric with black hole observations}},
               [\href{http://xxx.lanl.gov/abs/2002.12786}
{{\tt arXiv:2002.12786}}].




\bibitem{Sau:2020xau}
  S.~Sau, I.~Banerjee and S.~SenGupta,
{\it{Imprints of the Janis-Newman-Winicour spacetime on observations related to
shadow and accretion}},
  Phys.\ Rev.\ D {\bf 102}, no. 6, 064027 (2020)
                 [\href{http://xxx.lanl.gov/abs/2004.02840}
{{\tt arXiv:2004.02840}}].




 \bibitem{Belhaj:2020rdb}
  A.~Belhaj, M.~Benali, A.~El Balali, H.~El Moumni and S.~E.~Ennadifi,
 {\it{Deflection angle and shadow behaviors of quintessential black holes in
arbitrary dimensions}},
  Class.\ Quant.\ Grav.\  {\bf 37}, no. 21, 215004 (2020)
                   [\href{http://xxx.lanl.gov/abs/2006.01078}
{{\tt arXiv:2006.01078}}].




\bibitem{Kumar:2020yem}
  R.~Kumar, A.~Kumar and S.~G.~Ghosh,
{\it{Testing Rotating Regular Metrics as Candidates for Astrophysical Black
Holes}},
  Astrophys.\ J.\  {\bf 896}, no. 1, 89 (2020)
                     [\href{http://xxx.lanl.gov/abs/2006.09869}
{{\tt arXiv:2006.09869}}].




  \bibitem{Zeng:2020vsj}
  X.~X.~Zeng and H.~Q.~Zhang,
 {\it{Influence of quintessence dark energy on the shadow of black hole}},
  Eur.\ Phys.\ J.\ C {\bf 80}, no. 11, 1058
                       [\href{http://xxx.lanl.gov/abs/2007.06333}
{{\tt arXiv:2007.06333}}].


 \bibitem{Saurabh:2020zqg}
  Saurabh and K.~Jusufi,
 {\it{Imprints of Dark Matter on Black Hole Shadows using Spherical
Accretions}},
  [\href{http://xxx.lanl.gov/abs/2009.10599}
{{\tt arXiv:2009.10599}}].










\bibitem{Bambi:2019tjh}
C.~Bambi, K.~Freese, S.~Vagnozzi and L.~Visinelli,
{\it{Testing the rotational nature of the supermassive object M87* from the
circularity and size of its first image}},
Phys.\ Rev.\ D {\bf 100}, no. 4, 044057 (2019)
  [\href{http://xxx.lanl.gov/abs/1904.12983}
{{\tt arXiv:1904.12983}}].




\bibitem{Vagnozzi:2019apd}
  S.~Vagnozzi and L.~Visinelli,
{\it{Hunting for extra dimensions in the shadow of M87*}},
  Phys.\ Rev.\ D {\bf 100}, no. 2, 024020 (2019)
    [\href{http://xxx.lanl.gov/abs/1905.12421}
{{\tt arXiv:1905.12421}}].




 \bibitem{Haroon:2019new}
 S.~Haroon, K.~Jusufi and M.~Jamil,
 {\it{Shadow Images of a Rotating Dyonic Black Hole with a Global Monopole
Surrounded by Perfect Fluid}},
 Universe {\bf 6}, no. 2, 23 (2020)
     [\href{http://xxx.lanl.gov/abs/1904.00711}
{{\tt arXiv:1904.00711}}].


\bibitem{Shaikh:2019hbm}
  R.~Shaikh and P.~S.~Joshi,
{\it{Can we distinguish black holes from naked singularities by the images of
their accretion disks?}},
  JCAP {\bf 1910}, 064 (2019)
       [\href{http://xxx.lanl.gov/abs/1909.10322}
{{\tt arXiv:1909.10322}}].


\bibitem{Cunha:2019ikd}
  P.~V.~P.~Cunha, C.~A.~R.~Herdeiro and E.~Radu,
{\it{EHT constraint on the ultralight scalar hair of the M87 supermassive black
hole}},
  Universe {\bf 5}, no. 12, 220 (2019)
         [\href{http://xxx.lanl.gov/abs/1909.08039}
{{\tt arXiv:1909.08039}}].



\bibitem{Banerjee:2019nnj}
  I.~Banerjee, S.~Chakraborty and S.~SenGupta,
{\it{Silhouette of M87*: A New Window to Peek into the World of Hidden
Dimensions}},
  Phys.\ Rev.\ D {\bf 101}, no. 4, 041301 (2020)
          [\href{http://xxx.lanl.gov/abs/1909.09385}
{{\tt arXiv:1909.09385}}].



\bibitem{Feng:2019zzn}
  X.~H.~Feng and H.~Lu,
{\it{On the size of rotating black holes}},
  Eur.\ Phys.\ J.\ C {\bf 80}, no. 6, 551 (2020)
            [\href{http://xxx.lanl.gov/abs/1911.12368}
{{\tt arXiv:1911.12368}}].




\bibitem{Li:2019lsm}
  S.~F.~Yan, C.~Li, L.~Xue, X.~Ren, Y.~F.~Cai, D.~A.~Easson, Y.~F.~Yuan and
H.~Zhao,
{\it{Testing the equivalence principle via the shadow of black holes}},
  Phys.\ Rev.\ Res.\  {\bf 2}, no. 2, 023164 (2020)
             [\href{http://xxx.lanl.gov/abs/1912.12629}
{{\tt arXiv:1912.12629}}].




\bibitem{Allahyari:2019jqz}
  A.~Allahyari, M.~Khodadi, S.~Vagnozzi and D.~F.~Mota,
{\it{Magnetically charged black holes from non-linear electrodynamics and the
Event Horizon Telescope}},
  JCAP {\bf 2002}, 003 (2020)
               [\href{http://xxx.lanl.gov/abs/1912.08231}
{{\tt arXiv:1912.08231}}].


\bibitem{Rummel:2019ads}
  M.~Rummel and C.~P.~Burgess,
{\it{Constraining Fundamental Physics with the Event Horizon Telescope}},
  JCAP {\bf 2005}, 051 (2020)
                 [\href{http://xxx.lanl.gov/abs/2001.00041}
{{\tt arXiv:2001.00041}}].




\bibitem{Vagnozzi:2020quf}
  S.~Vagnozzi, C.~Bambi and L.~Visinelli,
{\it{Concerns regarding the use of black hole shadows as standard rulers}},
  Class.\ Quant.\ Grav.\  {\bf 37}, no. 8, 087001 (2020)
                   [\href{http://xxx.lanl.gov/abs/2001.02986}
{{\tt arXiv:2001.02986}}].



\bibitem{Khodadi:2020jij}
  M.~Khodadi, A.~Allahyari, S.~Vagnozzi and D.~F.~Mota,
{\it{Black holes with scalar hair in light of the Event Horizon Telescope}},
  JCAP {\bf 2009}, 026 (2020)
                     [\href{http://xxx.lanl.gov/abs/2005.05992}
{{\tt arXiv:2005.05992}}].



\bibitem{Chang:2020lmg}
  Z.~Chang and Q.~H.~Zhu,
{\it{Does the shape of the shadow of a black hole depend on motional status of
an observer?}},
  Phys.\ Rev.\ D {\bf 102}, no. 4, 044012 (2020)
                       [\href{http://xxx.lanl.gov/abs/2006.00685}
{{\tt arXiv:2006.00685}}].




\bibitem{Kruglov:2020tes}
  S.~I.~Kruglov,
{\it{The shadow of M87* black hole within rational nonlinear electrodynamics}},
  Mod.\ Phys.\ Lett.\ A {\bf 35}, no. 35, 2050291 (2020).

\bibitem{Ghosh:2020tdu}
  D.~Ghosh, A.~Thalapillil and F.~Ullah,
{\it{Astrophysical hints for magnetic black holes}},
                       [\href{http://xxx.lanl.gov/abs/2009.03363}
{{\tt arXiv:2009.03363}}].


\bibitem{Psaltis:2020lvx}
  D.~Psaltis {\it et al.} [Event Horizon Telescope Collaboration],
  {\it{Gravitational Test Beyond the First Post-Newtonian Order with the Shadow
of the M87 Black Hole}},
  Phys.\ Rev.\ Lett.\  {\bf 125}, no. 14, 141104 (2020)
    [\href{http://xxx.lanl.gov/abs/2010.01055}
{{\tt arXiv:2010.01055}}].



\bibitem{Jusufi:2019ltj}
  K.~Jusufi,
{\it{Quasinormal Modes of Black Holes Surrounded by Dark Matter and Their
Connection with the Shadow Radius}},
  Phys.\ Rev.\ D {\bf 101}, no. 8, 084055 (2020)
      [\href{http://xxx.lanl.gov/abs/1912.13320}
{{\tt arXiv:1912.13320}}].



\bibitem{Kumar:2019pjp}
  R.~Kumar, S.~G.~Ghosh and A.~Wang,
{\it{Shadow cast and deflection of light by charged rotating regular black
holes}},
  Phys.\ Rev.\ D {\bf 100}, no. 12, 124024 (2019)
        [\href{http://xxx.lanl.gov/abs/1912.05154}
{{\tt arXiv:1912.05154}}].



\bibitem{Konoplya:2020bxa}
  R.~A.~Konoplya and A.~F.~Zinhailo,
{\it{Quasinormal modes, stability and shadows of a black hole in the 4D
Einstein–Gauss–Bonnet gravity}},
  Eur.\ Phys.\ J.\ C {\bf 80}, no. 11, 1049 (2020)
          [\href{http://xxx.lanl.gov/abs/2003.01188}
{{\tt arXiv:2003.01188}}].



\bibitem{Liu:2020ola}
  C.~Liu, T.~Zhu, Q.~Wu, K.~Jusufi, M.~Jamil, M.~Azreg-Aïnou and A.~Wang,
{\it{Shadow and Quasinormal Modes of a Rotating Loop Quantum Black Hole}},
  Phys.\ Rev.\ D {\bf 101}, no. 8, 084001 (2020)
            [\href{http://xxx.lanl.gov/abs/2003.00477}
{{\tt arXiv:2003.00477}}].



\bibitem{Jusufi:2020dhz}
  K.~Jusufi,
{\it{Connection Between the Shadow Radius and Quasinormal Modes in Rotating
Spacetimes}},
  Phys.\ Rev.\ D {\bf 101}, no. 12, 124063 (2020)
              [\href{http://xxx.lanl.gov/abs/2004.04664}
{{\tt arXiv:2004.04664}}].




\bibitem{Jusufi:2020agr}
  K.~Jusufi, M.~Amir, M.~S.~Ali and S.~D.~Maharaj,
{\it{Quasinormal modes, shadow and greybody factors of 5D electrically charged
Bardeen black holes}},
  Phys.\ Rev.\ D {\bf 102}, no. 6, 064020 (2020)
               [\href{http://xxx.lanl.gov/abs/2005.11080}
{{\tt arXiv:2005.11080}}].


\bibitem{Jusufi:2020odz}
  K.~Jusufi, M.~Azreg-Aïnou, M.~Jamil, S.~W.~Wei, Q.~Wu and A.~Wang,
{\it{Quasinormal modes, quasiperiodic oscillations and shadow of rotating
regular black holes in non-minimally coupled Einstein-Yang-Mills theory}},
               [\href{http://xxx.lanl.gov/abs/2008.08450}
{{\tt arXiv:2008.08450}}].

\bibitem{Ghasemi-Nodehi:2020oiz}
  M.~Ghasemi-Nodehi, M.~Azreg-Aïnou, K.~Jusufi and M.~Jamil,
  {\it{Shadow, quasinormal modes, and quasiperiodic oscillations of rotating
Kaluza-Klein black holes}},
  Phys.\ Rev.\ D {\bf 102}, no. 10, 104032 (2020)
                 [\href{http://xxx.lanl.gov/abs/2011.02276}
{{\tt arXiv:2011.02276}}].



\bibitem{Olivares:2019dsc} 
H.~Olivares, O.~Porth, J.~Davelaar, E.~R.~Most, C.~M.~Fromm, Y.~Mizuno, Z.~Younsi and L.~Rezzolla,
{\it{Constrained transport and adaptive mesh refinement in the Black Hole Accretion Code}},
Astron.\ Astrophys.\  {\bf 629}, A61 (2019)
[\href{http://xxx.lanl.gov/abs/1906.10795}
{{ \tt arXiv:1906.10795}}].



\bibitem{White:2019wix} 
C.~J.~White,
{\it{Development and Application of Numerical Techniques for General-Relativistic Magnetohydrodynamics Simulations of Black Hole Accretion}},
[\href{http://xxx.lanl.gov/abs/1906.09708}
{{\tt arXiv:1906.09708}}].

\bibitem{Nathanail:2020wap} 
A.~Nathanail, C.~M.~Fromm, O.~Porth, H.~Olivares, Z.~Younsi, Y.~Mizuno and L.~Rezzolla,
{\it{Plasmoid formation in global GRMHD simulations and AGN flares}},
Mon.\ Not.\ Roy.\ Astron.\ Soc.\  {\bf 495}, no. 2, 1549 (2020)
[\href{http://xxx.lanl.gov/abs/20002.01777}
{{\tt arXiv:2002.01777}}].

\bibitem{Bronzwaer:2020vix} 
T.~Bronzwaer, J.~Davelaar, Z.~Younsi, M.~Mościbrodzka, H.~Olivares, Y.~Mizuno, J.~Vos and H.~Falcke,
{\it{Visibility of Black Hole Shadows in Low-luminosity AGN}},
Mon.\ Not.\ Roy.\ Astron.\ Soc.\  {\bf 501}, no. 4, 4722 (2021)
[\href{http://xxx.lanl.gov/abs/2011.00069}
{{\tt arXiv:2011.00069}}].


\bibitem{Cruz-Osorio:2021gnz} 
A.~Cruz-Osorio, S.~Gimeno-Soler, J.~A.~Font, M.~De Laurentis and S.~Mendoza,
{\it{Magnetized discs and photon rings around Yukawa-like black holes}},
[\href{http://xxx.lanl.gov/abs/2102.10150}
{{arXiv:2102.10150}}].



\bibitem{Mattingly:2005re}
  D.~Mattingly,
{\it{Modern tests of Lorentz invariance}},
  Living Rev.\ Rel.\  {\bf 8}, 5 (2005)
                   [\href{http://xxx.lanl.gov/abs/gr-qc/0502097}
{{\tt arXiv:gr-qc/0502097}}].




\bibitem{Will:2005va}
C.~M.~Will,
{\it{The Confrontation between general relativity and experiment}},
Living Rev.\ Rel.\  {\bf 9}, 3 (2006)
                   [\href{http://xxx.lanl.gov/abs/gr-qc/0510072}
{{\tt arXiv:gr-qc/0510072}}].



\bibitem{Liberati:2015dja}
  S.~Liberati,
{\it{Lorentz symmetry breaking: phenomenology and constraints}},
  J.\ Phys.\ Conf.\ Ser.\  {\bf 631}, no. 1, 012011 (2015).

\bibitem{Colladay:1998fq}
  D.~Colladay and V.~A.~Kostelecky,
  {\it{Lorentz violating extension of the standard model}},
  Phys.\ Rev.\ D {\bf 58}, 116002 (1998)
                     [\href{http://xxx.lanl.gov/abs/hep-ph/9809521}
{{\tt arXiv:hep-ph/9809521}}].




\bibitem{Kostelecky:2003fs}
  V.~A.~Kostelecky,
 {\it{Gravity, Lorentz violation, and the standard model}},
  Phys.\ Rev.\ D {\bf 69}, 105009 (2004)
                       [\href{http://xxx.lanl.gov/abs/hep-th/0312310}
{{\tt arXiv:hep-th/0312310}}].



\bibitem{Jacobson:2000xp}
T.~Jacobson and D.~Mattingly,
{\it{Gravity with a dynamical preferred frame}},
Phys.\ Rev.\ D {\bf 64}, 024028 (2001)
                       [\href{http://xxx.lanl.gov/abs/gr-qc/0007031}
{{\tt arXiv:gr-qc/0007031}}].


\bibitem{Eling:2004dk}
C.~Eling, T.~Jacobson and D.~Mattingly,
{\it{Einstein-\AE ther theory}},
                       [\href{http://xxx.lanl.gov/abs/gr-qc/0410001}
{{\tt arXiv:gr-qc/0410001}}].


\bibitem{Jacobson:2008aj}
T.~Jacobson,
{\it{Einstein-\AE ther gravity: A Status report}},
PoS QG {\bf -PH}, 020 (2007)
                       [\href{http://xxx.lanl.gov/abs/0801.1547}
{{\tt arXiv:0801.1547}}].





\bibitem{Foster:2005dk}
  B.~Z.~Foster and T.~Jacobson,
{\it{Post-Newtonian parameters and constraints on Einstein-\AE ther theory}},
  Phys.\ Rev.\ D {\bf 73}, 064015 (2006)
                      [\href{http://xxx.lanl.gov/abs/gr-qc/0509083}
{{\tt arXiv:gr-qc/0509083}}].




\bibitem{Elliott:2005va}
  J.~W.~Elliott, G.~D.~Moore and H.~Stoica,
{\it{Constraining the new Aether: Gravitational Cerenkov radiation}},
  JHEP {\bf 0508}, 066 (2005)
                        [\href{http://xxx.lanl.gov/abs/hep-ph/0505211}
{{\tt arXiv:hep-ph/0505211}}].



\bibitem{Li:2007vz}
  B.~Li, D.~Fonseca Mota and J.~D.~Barrow,
  {\it{Detecting a Lorentz-Violating Field in Cosmology}},
  Phys.\ Rev.\ D {\bf 77}, 024032 (2008)
     [\href{http://xxx.lanl.gov/abs/0709.4581}
{{\tt arXiv:0709.4581}}].



\bibitem{Yagi:2013ava}
  K.~Yagi, D.~Blas, E.~Barausse and N.~Yunes,
  {\it{Constraints on Einstein-Æther theory and Hořava gravity from binary
pulsar observations}},
  Phys.\ Rev.\ D {\bf 89}, no. 8, 084067 (2014)
  Erratum: [Phys.\ Rev.\ D {\bf 90}, no. 6, 069902 (2014)]
  Erratum: [Phys.\ Rev.\ D {\bf 90}, no. 6, 069901 (2014)]
       [\href{http://xxx.lanl.gov/abs/1311.7144}
{{\tt arXiv:1311.7144}}].



\bibitem{Horava:2009uw}
  P.~Horava,
  {\it{Quantum Gravity at a Lifshitz Point}},
  Phys.\ Rev.\ D {\bf 79}, 084008 (2009)
                         [\href{http://xxx.lanl.gov/abs/0901.3775}
{{\tt arXiv:0901.3775}}].



\bibitem{Mukohyama:2010xz}
  S.~Mukohyama,
  {\it{Horava-Lifshitz Cosmology: A Review}},
  Class.\ Quant.\ Grav.\  {\bf 27}, 223101 (2010)
                           [\href{http://xxx.lanl.gov/abs/1007.5199}
{{\tt arXiv:1007.5199}}].




\bibitem{Wang:2017brl}
A.~Wang,
{\it{Horava gravity at a Lifshitz point: A progress report}},
Int.\ J.\ Mod.\ Phys.\ D {\bf 26}, 1730014 (2017),
              [\href{http://xxx.lanl.gov/abs/1701.06087}
{{\tt arXiv:1701.06087}}].



\bibitem{Chamseddine:2013kea}
  A.~H.~Chamseddine and V.~Mukhanov,
{\it{Mimetic Dark Matter}},
  JHEP {\bf 1311}, 135 (2013)
                [\href{http://xxx.lanl.gov/abs/1308.5410}
{{\tt arXiv:1308.5410}}].



\bibitem{Chamseddine:2014vna}
  A.~H.~Chamseddine, V.~Mukhanov and A.~Vikman,
{\it{Cosmology with Mimetic Matter}},
  JCAP {\bf 1406}, 017 (2014)
                  [\href{http://xxx.lanl.gov/abs/1403.3961}
{{\tt arXiv:1403.3961}}].


\bibitem{Basilakos:2013hua}
S.~Basilakos, A.~P.~Kouretsis, E.~N.~Saridakis and P.~Stavrinos,
{\it{Resembling dark energy and modified gravity with Finsler-Randers
cosmology}},
Phys. Rev. D \textbf{88}, 123510 (2013)
                  [\href{http://xxx.lanl.gov/abs/1311.5915}
{{\tt arXiv:1311.5915}}].



\bibitem{Ikeda:2019ckp}
S.~Ikeda, E.~N.~Saridakis, P.~C.~Stavrinos and A.~Triantafyllopoulos,
{\it{Cosmology of Lorentz fiber-bundle induced scalar-tensor theories}},
Phys. Rev. D \textbf{100}, no.12, 124035 (2019)
                  [\href{http://xxx.lanl.gov/abs/1907.10950}
{{\tt arXiv:1907.10950}}].





\bibitem{Sebastiani:2016ras}
  L.~Sebastiani, S.~Vagnozzi and R.~Myrzakulov,
  {\it{Mimetic gravity: a review of recent developments and applications to
cosmology and astrophysics}},
  Adv.\ High Energy Phys.\  {\bf 2017}, 3156915 (2017)
                    [\href{http://xxx.lanl.gov/abs/1612.08661}
{{\tt arXiv:1612.08661}}].



\bibitem{Oost:2018tcv}
J.~Oost, S.~Mukohyama and A.~Wang,
{\it{Constraints on Einstein-\AE ther theory after GW170817}},
Phys.\ Rev.\ D {\bf 97}, no. 12, 124023 (2018)
                    [\href{http://xxx.lanl.gov/abs/1802.04303}
{{\tt arXiv:1802.04303}}].




\bibitem{Eling:2006ec}
C.~Eling and T.~Jacobson,
{\it{Black Holes in Einstein-\AE ther Theory}},
Class.\ Quant.\ Grav.\  {\bf 23}, 5643 (2006)
Erratum: [Class.\ Quant.\ Grav.\  {\bf 27}, 049802 (2010)]
       [\href{http://xxx.lanl.gov/abs/gr-qc/0604088}
{{\tt arXiv:gr-qc/0604088}}].




\bibitem{Barausse:2011pu}
E.~Barausse, T.~Jacobson and T.~P.~Sotiriou,
{\it{Black holes in Einstein-\AE ther and Horava-Lifshitz gravity}},
Phys.\ Rev.\ D {\bf 83}, 124043 (2011)
       [\href{http://xxx.lanl.gov/abs/1104.2889}
{{\tt arXiv:1104.2889}}].




\bibitem{Ding:2015kba}
C.~Ding, A.~Wang and X.~Wang,
{\it{Charged Einstein-\AE ther black holes and Smarr formula}},
Phys.\ Rev.\ D {\bf 92}, no. 8, 084055 (2015)
       [\href{http://xxx.lanl.gov/abs/1507.06618}
{{\tt arXiv:1507.06618}}].


\bibitem{Zhang:2020too}
C.~Zhang, X.~Zhao, K.~Lin, S.~Zhang, W.~Zhao and A.~Wang,
{\it{Spherically symmetric static black holes in Einstein-aether theory}},
Phys. Rev. D \textbf{102}, no.6, 064043 (2020)
       [\href{http://xxx.lanl.gov/abs/2004.06155}
{{\tt arXiv:2004.06155}}].



\bibitem{Psaltis:2014mca}
D.~Psaltis, F.~Ozel, C.~K.~Chan and D.~P.~Marrone,
{\it{A General Relativistic Null Hypothesis Test with Event Horizon Telescope
Observations of the black-hole shadow in Sgr A*}},
Astrophys.\ J.\  {\bf 814}, no. 2, 115 (2015)
       [\href{http://xxx.lanl.gov/abs/1411.1454}
{{\tt arXiv:1411.1454}}].



\bibitem{Carroll:2004ai}
  S.~M.~Carroll and E.~A.~Lim,
{\it{Lorentz-violating vector fields slow the universe down}},
  Phys.\ Rev.\ D {\bf 70}, 123525 (2004)
         [\href{http://xxx.lanl.gov/abs/hep-th/0407149}
{{\tt arXiv:hep-th/0407149}}].



\bibitem{Berglund:2012bu}
P.~Berglund, J.~Bhattacharyya and D.~Mattingly,
{\it{Mechanics of universal horizons}},
Phys.\ Rev.\ D {\bf 85}, 124019 (2012)
       [\href{http://xxx.lanl.gov/abs/1202.4497}
{{\tt arXiv:1202.4497}}].




\bibitem{Jacobson:2007fh}
 T.~Jacobson,
{\it{Einstein-\AE ther gravity: Theory and observational constraints}},
         [\href{http://xxx.lanl.gov/abs/0711.3822}
{{\tt arXiv:0711.3822}}].




\bibitem{Gorji:2019rlm}
M.~A.~Gorji, A.~Allahyari, M.~Khodadi and H.~Firouzjahi,
{\it{Mimetic black holes}},
Phys.\ Rev.\ D {\bf 101}, no. 12, 124060 (2020)
       [\href{http://xxx.lanl.gov/abs/1912.04636}
{{\tt arXiv:1912.04636}}].


\bibitem{Hartle:1968si}
J.~B.~Hartle and K.~S.~Thorne,
{\it{Slowly Rotating Relativistic Stars. II. Models for Neutron Stars and
Supermassive Stars}},
Astrophys.\ J.\  {\bf 153}, 807 (1968).

\bibitem{Barausse:2015frm}
E.~Barausse, T.~P.~Sotiriou and I.~Vega,
{\it{Slowly rotating black holes in Einstein-æther theory}},
Phys.\ Rev.\ D {\bf 93}, no. 4, 044044 (2016)
         [\href{http://xxx.lanl.gov/abs/1512.05894}
{{\tt arXiv:1512.05894}}].




\bibitem{Barausse:2013nwa}
 E.~Barausse and T.~P.~Sotiriou,
  {\it{Black holes in Lorentz-violating gravity theories}},
  Class.\ Quant.\ Grav.\  {\bf 30}, 244010 (2013)
           [\href{http://xxx.lanl.gov/abs/1307.3359}
{{\tt arXiv:1307.3359}}].



\bibitem{Wang:2012nv}
  A.~Wang,
  {\it{Stationary axisymmetric and slowly rotating spacetimes in
Horava-lifshitz gravity}},
  Phys.\ Rev.\ Lett.\  {\bf 110}, 091101 (2013)
             [\href{http://xxx.lanl.gov/abs/1212.1876}
{{\tt arXiv:1212.1876}}].



\bibitem{Reynolds:1996fh} 
C.~S.~Reynolds, A.~C.~Fabian, A.~Celotti and M.~J.~Rees,
{\it{The matter content of the jet in m87: evidence for an electron-positron jet}},
Mon.\ Not.\ Roy.\ Astron.\ Soc.\  {\bf 283}, 873 (1996)
[\href{http://xxx.lanl.gov/abs/9603140}
{{\tt astro-ph/9603140}}].


\bibitem{deGasperin:2012id} 
F.~de Gasperin {\it et al.},
{\it{M87 at metre wavelengths: the LOFAR picture}},
Astron.\ Astrophys.\  {\bf 547}, A56 (2012)
[\href{http://xxx.lanl.gov/abs/1210.1346}
{{\tt arXiv:1210.1346}}].

\bibitem{Broderick:2015tda} 
A.~E.~Broderick, R.~Narayan, J.~Kormendy, E.~S.~Perlman, M.~J.~Rieke and S.~S.~Doeleman,
{\it{The Event Horizon of M87}},
Astrophys.\ J.\  {\bf 805}, no. 2, 179 (2015)
[\href{http://xxx.lanl.gov/abs/1503.03873}
{{\tt arXiv:1503.03873}}].


\bibitem{Stawarz:2006bk} 
L.~Stawarz, F.~Aharonian, J.~Kataoka, M.~Ostrowski, A.~Siemiginowska and M.~Sikora,
{\it{Dynamics and high energy emission of the flaring hst-1 knot in the m 87 jet}},
Mon.\ Not.\ Roy.\ Astron.\ Soc.\  {\bf 370}, 981 (2006)
[\href{http://xxx.lanl.gov/abs/0602220}
{{\tt arXiv:astro-ph/0602220}}].






\bibitem{Reynolds:2020jwt}
  C.~S.~Reynolds,
  {\it{Observational Constraints on Black Hole Spin}},
               [\href{http://xxx.lanl.gov/abs/2011.08948}
{{\tt arXiv:2011.08948}}].


\bibitem{Reynolds:2013qqa}
C.~S.~Reynolds,
{\it{Measuring Black Hole Spin using X-ray Reflection Spectroscopy}},
Space Sci.\ Rev.\  {\bf 183}, no. 1-4, 277 (2014)
[\href{http://xxx.lanl.gov/abs/1212.1876}
{{\tt arXiv:1212.1876}}].


\bibitem{Volonteri:2004cf}
M.~Volonteri, P.~Madau, E.~Quataert and M.~J.~Rees,
{\it{The Distribution and cosmic evolution of massive black hole spins}},
Astrophys.\ J.\  {\bf 620}, 69 (2005)
[\href{http://xxx.lanl.gov/abs/astro-ph/0410342}
{{\tt arXiv:astro-ph/0410342}}].



\bibitem{Sesana:2014bea}
A.~Sesana, E.~Barausse, M.~Dotti and E.~M.~Rossi,
{\it{Linking the spin evolution of massive black holes to galaxy kinematics}},
Astrophys.\ J.\  {\bf 794}, 104 (2014)
[\href{http://xxx.lanl.gov/abs/1402.7088}
{{\tt arXiv:1402.7088}}].



\bibitem{Zhang:2019ekx}
  X.~Zhang and Y.~Lu,
  {\it{On Constraining the Growth History of Massive Black Holes via Their
Distribution on the Spin–Mass Plane}},
  Astrophys.\ J.\  {\bf 873}, no. 2, 101 (2019)
                     [\href{http://xxx.lanl.gov/abs/1902.07056}
{{\tt arXiv:1902.07056}}].




\bibitem{Bustamante:2019sjz}
  S.~Bustamante and V.~Springel,
{\it{Spin evolution and feedback of supermassive black holes in cosmological
simulations}},
  Mon.\ Not.\ Roy.\ Astron.\ Soc.\  {\bf 490}, no. 3, 4133 (2019)
  [\href{http://xxx.lanl.gov/abs/1902.04651}
{{\tt arXiv:1902.04651}}].

\end{thebibliography}
\end{document}